\documentclass[aps,prb,twocolumn,showpacs,superscriptaddress,groupedaddress,floatfix]{revtex4-1}
\usepackage{latexsym}
\usepackage{graphicx}
\usepackage{dcolumn}
\usepackage{bm}
\usepackage{amssymb}
\usepackage{amsmath}
\usepackage[space]{grffile}
\usepackage{color}

\usepackage{subfig}

\begin{document}

\newcommand{\niceref}[1] {Eq.~(\ref{#1})}
\newcommand{\fullref}[1] {Equation~(\ref{#1})}
\title{Time-resolved spectral density of interacting fermions following a quench to a superconducting critical point}

\date{\today}

\begin{abstract}
Results are presented for the time evolution of fermions initially in a non-zero temperature normal phase, following the switch on of an 
attractive interaction.  The dynamics are studied in the disordered phase close to the critical point, where the superfluid fluctuations 
are large. The analysis is conducted 
within a two-particle irreducible, large $N$ approximation.  The system is considered from the perspective of critical quenches where it is 
shown that the fluctuations follow universal model A dynamics. A signature of this universality is found in a singular correction to the 
fermion lifetime, given by a scaling form $t^{(3-d)/2}S_d(\varepsilon^2 t)$, where $d$ is the spatial dimension, $t$ is the time 
since the quench, and $\varepsilon$ is the 
fermion energy. The singular behavior of the spectral density is interpreted as arising due to incoherent Andreev reflections 
off superfluid fluctuations.
\end{abstract}

\author{Yonah Lemonik}
\author{Aditi Mitra}

\affiliation{Department of Physics, New York University, 726 Broadway, New York, NY, 10003, USA}
\maketitle
\section{Introduction}

Experiments involving pump-probe spectroscopy of solid-state systems~\cite{Fausti11,Smallwood12,Smallwood14,Beck13,Mitrano15} 
as well as dynamics of cold-atomic gases~\cite{Regal04,Zwierlein04,Bloch08,Bloch12,Navon16} have opened up an entirely new
temporal regime for probing correlated systems. A strong disturbance at
an initial time, either by a pump beam of a few femto-second duration, or by explicit changes to the lattice and interaction parameters 
for atoms confined in an optical lattice, places the system in a highly nonequilibrium state. Following this,
the dynamics can be probed with pico-second resolution for solid state systems, and milli-second resolution for
cold atoms. At these short time-scales, the system is far from thermal equilibrium, and has memory of the initial pulse.

For generic systems, with only a finite number of conservation laws such as 
energy and particle number, relaxation to thermal equilibrium is expected to be 
fast~\cite{Juchem04,Werner09,Santos10,Tavora14}.
However the relaxation may still possess rich dynamics when tuned near a critical point. As in equilibrium, observables will have a singular dependence on the detuning 
from the critical point. 
Such singular behavior in quench dynamics has already been identified for bosonic systems coupled to
a bath~\cite{Janssen1988,Huse89,Gambassi05,Gagel2014,Gagel15}, and also for isolated bosonic 
systems~\cite{Sondhi2013,Chiocchetta2015,Maraga2015,Oberthaler15,MitGam16,Lemonik16,Gasenzer17,Marino17}. In this paper, we complement this study
with that for an isolated fermionic system.

We consider a gas of fermions in a lattice without disorder at finite temperature, where the fermions have an attractive interaction. 
This can be realized as a gas of cold atoms in an optical lattice. In equilibrium, there is a phase transition separating superfluid and normal (non-superfluid) 
phases.  We study this phase transition as a quench process. In particular we study the dynamics when the system is initially in the normal phase, 
but where the interaction is suddenly increased so the system is  close to the phase transition.

Our results may be understood as follows. The distance from the critical point is captured by the superconducting fluctuation 
$D(q,t)\equiv \langle \Delta^{\dagger}(q,t)\Delta(q,t)\rangle$, where $\Delta^\dagger(q,0)$ creates a Cooper pair at momentum $q$.  
In the initial system, deep in the normal phase, $D(q)$ is small and non-singular  as a function of $q$. 
At the finite temperature equilibrium critical point, we expect $D(q,\infty) \sim q^{-2}$ as $q\rightarrow 0$. 
When a large interaction is turned on in the normal state, 
the system must interpolate between these two limits as a function of time. Since in a diffusive system information can only be 
communicated a distance 
$\propto\sqrt{t}$ in a time $t$ after the quench, we should expect a scale $\sim t^{-1/2}$ to function as the cutoff on the 
divergence of $D(q,t)$. Therefore the long wavelength fluctuations should show strong dependence on $t$.
\begin{figure}
\includegraphics[width = .95\columnwidth]{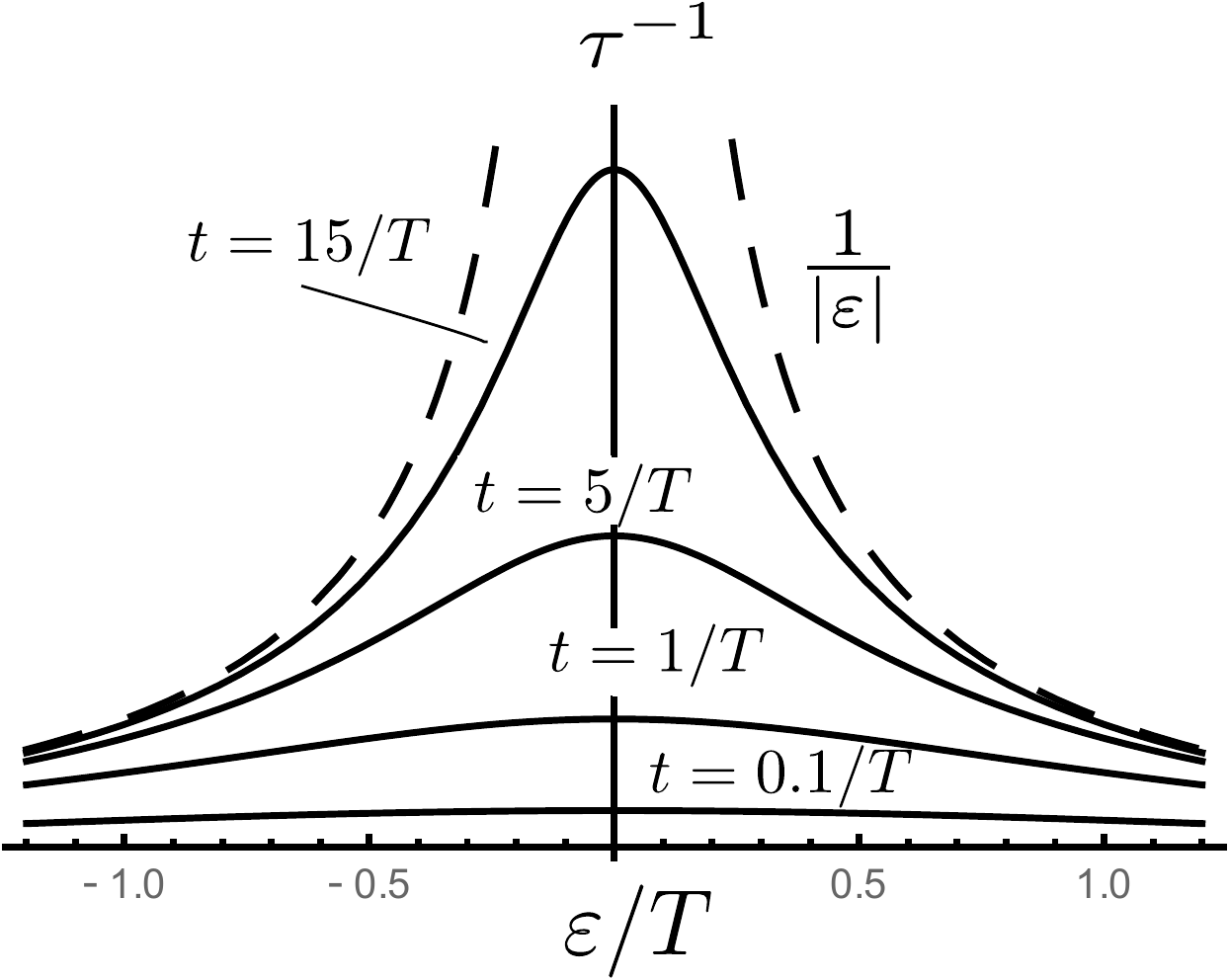}
\raisebox{-180pt}{
\includegraphics[width = .95\columnwidth]{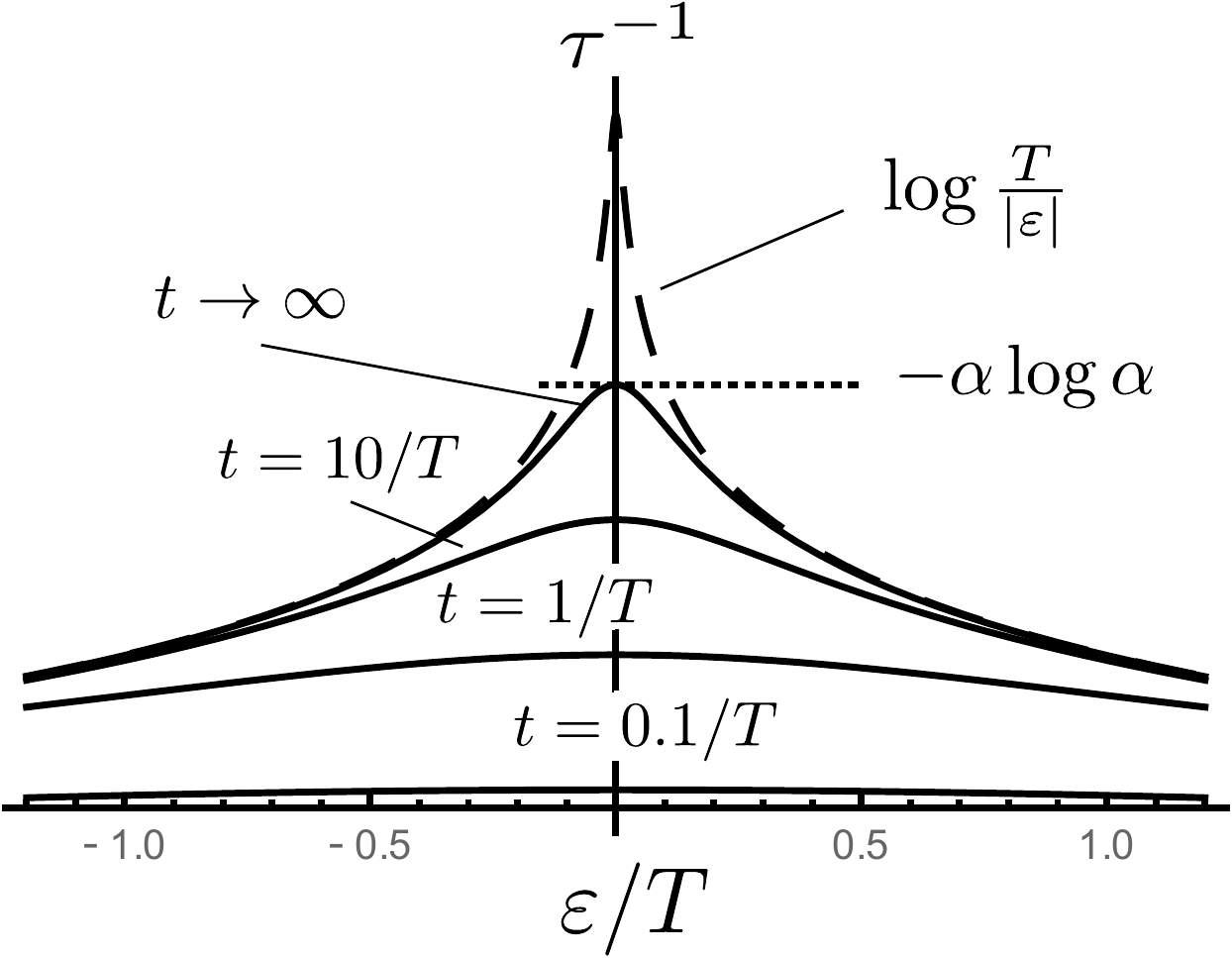}
}
\caption{
Predictions for the fermion lifetime $\tau^{-1}$ in $d=2$ (top panel) and $d=3$ (bottom panel), given in units of $\alpha T$,
where $T$ is the temperature of the electrons before the quench, and $\alpha$ is a material dependent dimensionless parameter.  
The lifetime is plotted as a function of fermion energy $\varepsilon$ divided by the temperature $T$, for several different times. 
The envelope giving the decay of the tails is indicated with dashed lines.  While the long time behavior of the lifetime is unclear in $d=2$, 
in $d=3$ it saturates to a fixed curve, sketched here schematically for $\alpha/4 \pi = 0.1$. \label{fig:summary}}
\end{figure}

We propose to detect these fluctuations through their effect on the spectral properties of the fermions, in the vein of fluctuation superconductivity. 
If we imagine a fixed background of superfluid fluctuations, the fermions can be thought of as Andreev reflecting off of this background. 
As the system is disordered the Andreev reflection is incoherent, and no gap is opened in the fermionic spectrum. 
However, the Andreev reflection is an energy conserving process for fermions at the chemical potential. This then contributes a decay channel 
for fermions at the Fermi surface which we can estimate by Fermi's golden rule as $\int d^{d-1}{q} D(q,t)$ (where the $d-1$ dimensional integral 
corresponds to the Fermi surface of a $d$ dimensional system). 
In $d=2,3$ this integral is singular as $q\rightarrow0, t\rightarrow\infty$, and we obtain a singular correction which goes like $\sqrt{t}$ in $d=2$ 
and $\log t$ in $d=3$. This behavior is summarized in Fig.~\ref{fig:summary}.

Our analysis is in a regime complementary to studies such as~\cite{Yuzbashyan15,Foster15,Foster14} 
where the initial state was already superconducting to begin with, and the dynamics of the superconducting order-parameter
under an interaction quench was studied within a mean field approximation. It is also complementary to studies 
such as~\cite{Sentef16,Knap16,Dehghani17,Kennes17}
where the initial state was in the normal phase, and the external perturbation puts the system deep in the ordered phase where the dynamics 
were again studied in mean field.  
In contrast, we study the behavior when the system is always in the disordered phase and the mean field behavior is trivial. 
To controllably go beyond mean-field calculation we perform a $1/N$ expansion, where $N$ is an additional orbital degree of freedom for the fermions.

The paper is organized follows. In Section~\ref{model} we present the model, outline the approximations, and
introduce an auxiliary or Hubbard-Stratonovich field that represents the Cooper pair fluctuations or Cooperons.
In Section~\ref{sec:pert} the equations of motions are analyzed assuming the Cooperons are non-interacting, an assumption valid at short times. 
Further the effect of the fluctuations on the fermion lifetime is calculated. 
In Section~\ref{sec:nonpert}, the longer time behavior is considered. This is done by mapping the dynamics to model-A~\cite{HH77}, 
and solving the self-consistent equation for the self-energy. We conclude in section~\ref{conclu}. The  
mapping to model A dynamics is 
outlined in Appendices \ref{ap1} and \ref{ap2}, where App.~\ref{ap2} includes a perturbative estimate of the parameters of
Model A. The implication of model A on Cooperon dynamics is relegated to App.~\ref{ap3}. 

\section{Model}\label{model}
We study a quench where the initial Hamiltonian is that of free fermions,
\begin{eqnarray}
&&H_i=\sum_{k,\sigma=\uparrow,\downarrow,\tau=1\ldots N}\epsilon_k c_{k\sigma \tau}^{\dagger}c_{k\sigma \tau}.
\end{eqnarray}
Above $k$ is the momentum, $\sigma=\uparrow,\downarrow$ denotes the spin, and $\tau$ is an orbital quantum
number that takes $N$ values.
We consider the initial state to be the ground state of $H_i$ at non-zero temperature $T$, and chemical potential $\mu$.
The time-evolution from $t>0$ is in the presence of a weak pairing interaction $u$. We write the quartic pairing interaction
in terms of pair operators $\Delta_q$ such that,
\begin{eqnarray}
&&H_f = H_i + \frac{u}{N} \sum_{q}\Delta^\dagger_{q}\Delta_{q},\nonumber \\
&&\Delta_{q} = \sum_{k\tau} c_{k,\uparrow,\tau}c_{-k+q,\downarrow,\tau};
\Delta^\dagger_{q} = \sum_{k,\tau} c^{\dagger}_{-k+q,\downarrow,\tau}c_{k\uparrow \tau}^{\dagger}.\label{Hf}
\end{eqnarray}
The Hamiltonian above assumes contact interaction, so that only fermions
with opposite spin quantum numbers scatter off of each other.
In the superfluid phase $\langle \Delta_q\rangle \neq 0$. In this paper, since we are always in the normal phase,
$\langle \Delta_q\rangle=0$. 

In the strict $N\rightarrow\infty$ limit, the fermion number at each momentum $k$ is conserved. Thus
the system is fully integrable, and fails to thermalize. We do not work in this limit and consider finite $1/N$ corrections to the behavior. 
In particular this allows us to study the back-reaction of the fluctuating Cooper pairs on the fermionic gas.

We use a two-particle irreducible ($2PI$) formalism~\cite{CornwallJackiwTomboulis,BergesRev} to obtain the equations of motions.  
The main ingredient is the sum of $2PI$ diagrams $\Gamma'[G]$, which is a functional of the fermions Greens functions $G$,
\begin{figure}
\begin{flalign*}
&{\raisebox{20pt}{a)}}\hspace{-3pt}\includegraphics[scale=.5]{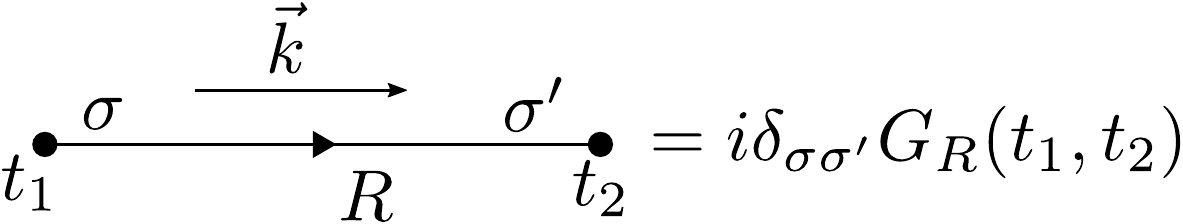}&\\
&{\raisebox{20pt}{b)}}\hspace{-3pt}\includegraphics[scale=.5]{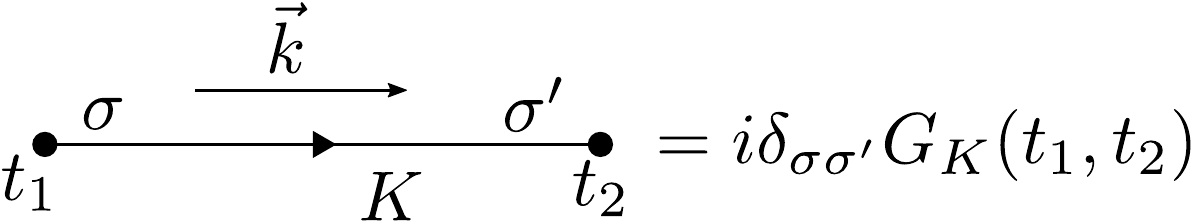}\\
&{\raisebox{20pt}{c)}}\hspace{-3pt}\includegraphics[scale=.5]{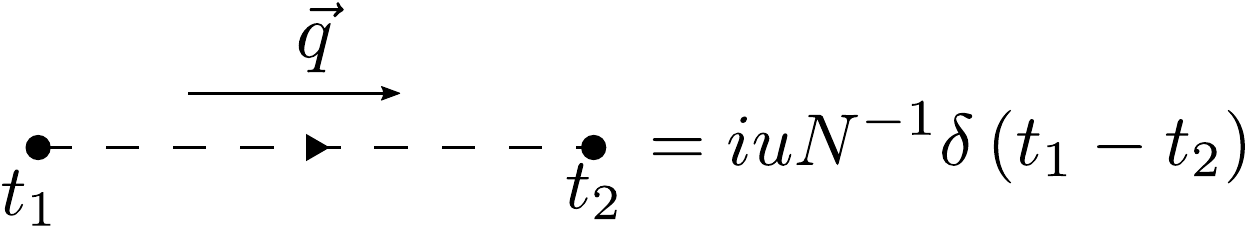}\\
&{\raisebox{35pt}{d)}}\hspace{-3pt}\includegraphics[scale=.5]{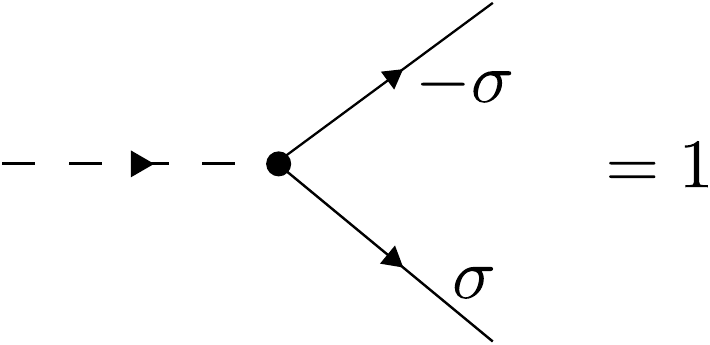}
\end{flalign*}
\caption{Elements of the Feynman diagrams: a) Retarded fermion Green's function. b) Keldysh Green's function. c) Fermion-fermion interaction. d) Interaction vertex. We suppress the orbital fermion index. \label{fig:FeynDefs}}
\end{figure}

\begin{figure}
\begin{flalign*}
&{\raisebox{135pt}{a)}}\hspace{-3pt}\includegraphics[scale=.45]{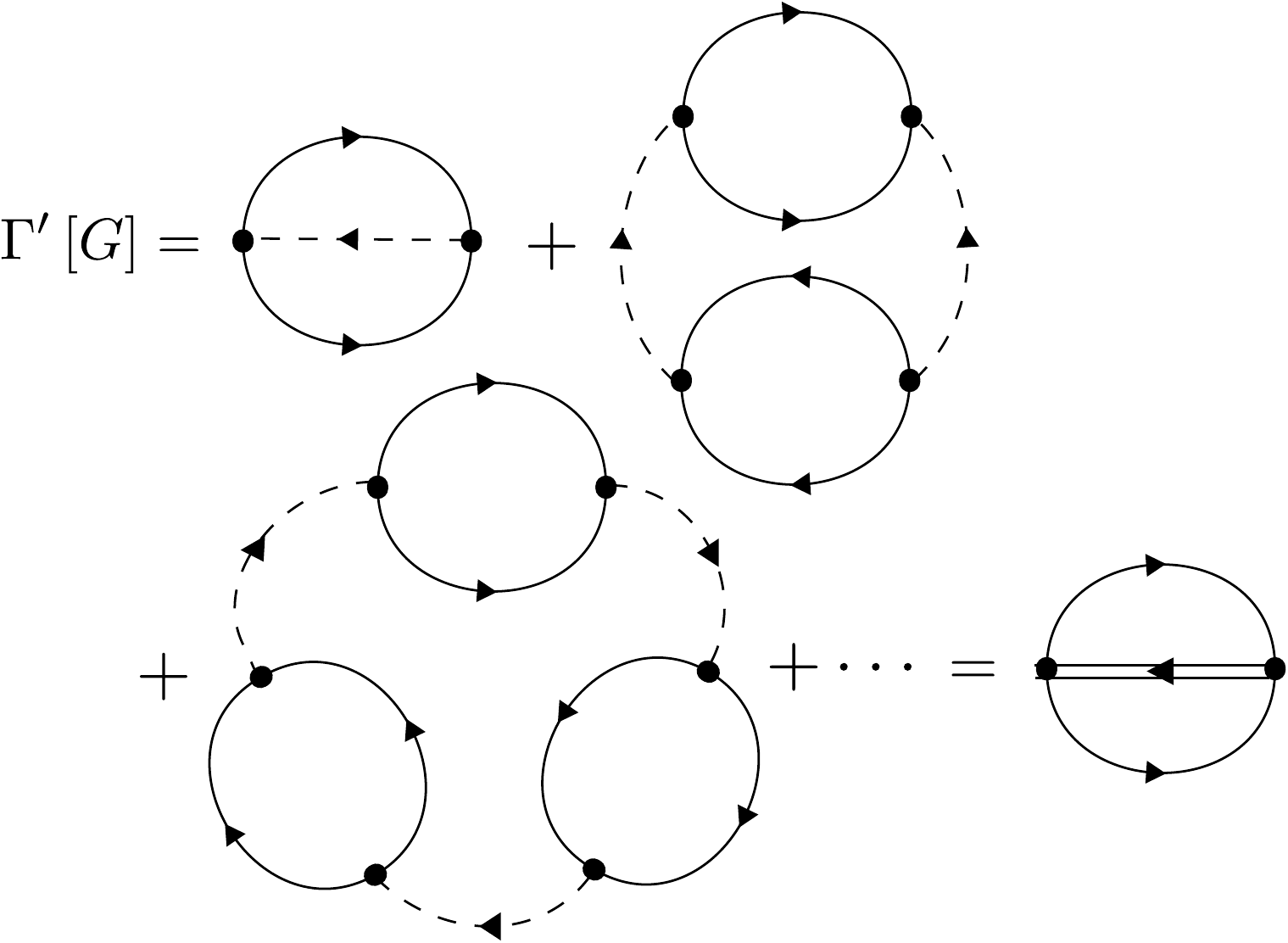}&\\
&{\raisebox{65pt}{b)}}\hspace{-1pt}\includegraphics[scale=.45]{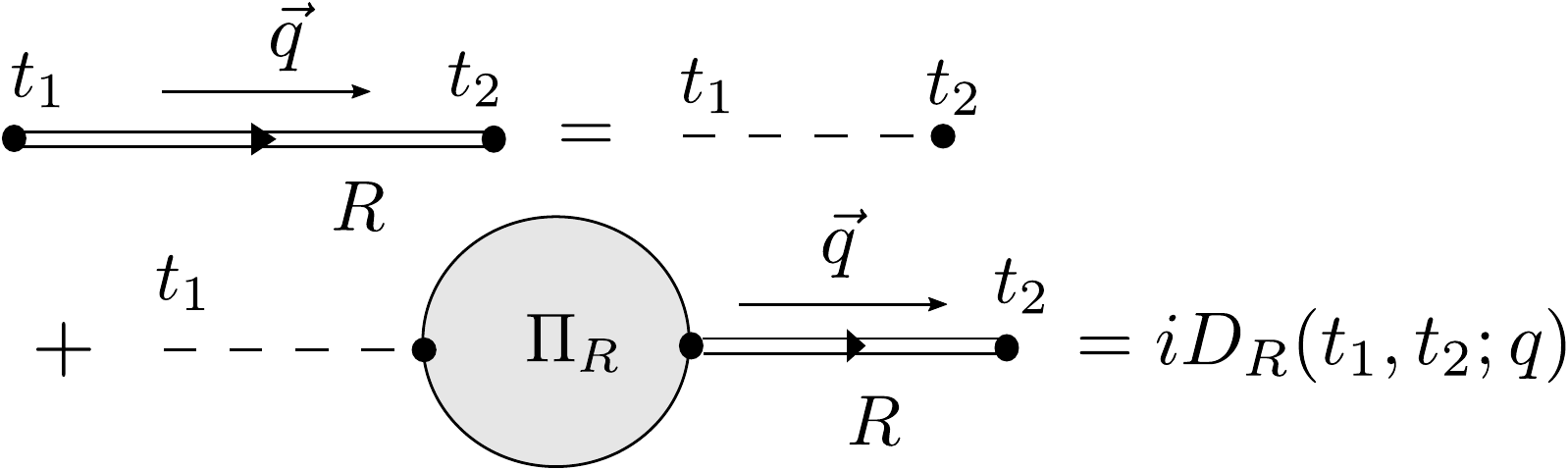}\\
&{\raisebox{55pt}{c)}}\hspace{-1pt}\includegraphics[scale=.45]{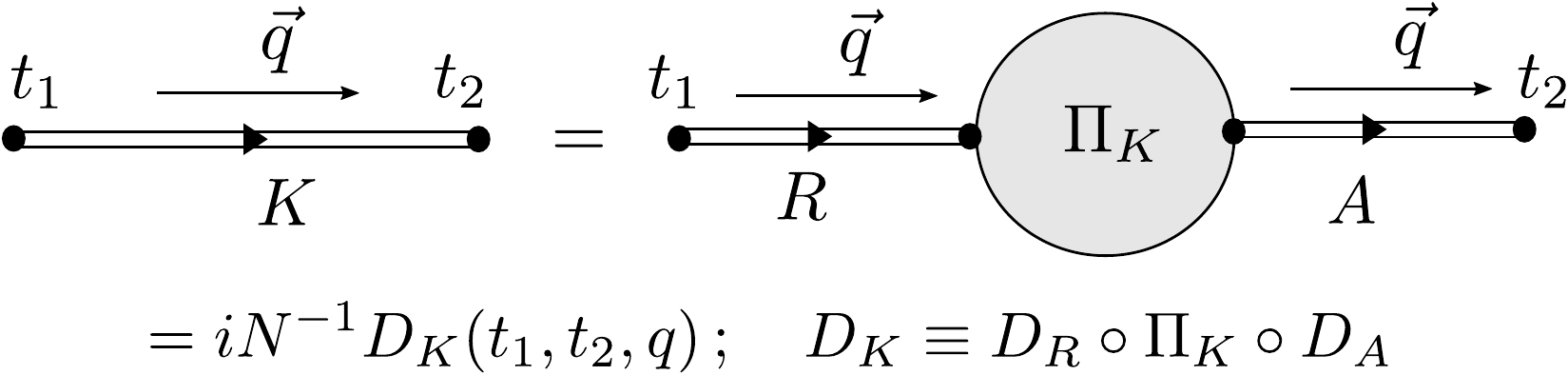}\\
&{\raisebox{35pt}{d)}}\hspace{-3pt}\includegraphics[scale=.45]{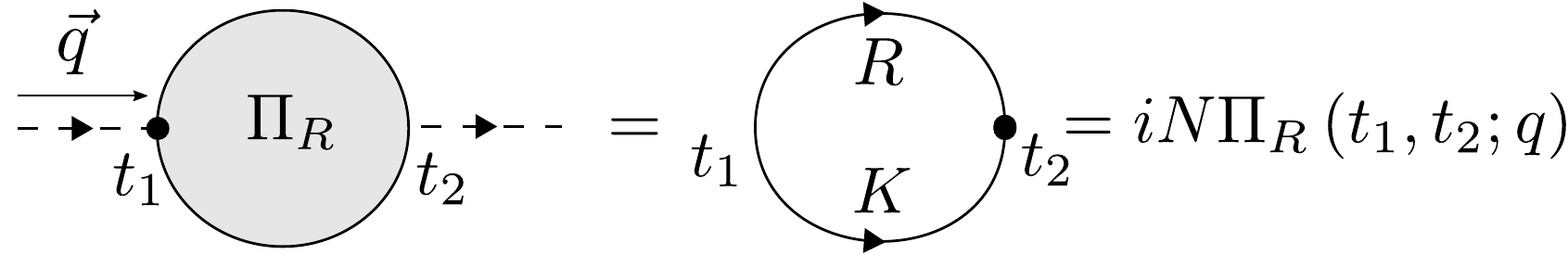}\\
&{\raisebox{78pt}{e)}}\hspace{-3pt}\includegraphics[scale=.45]{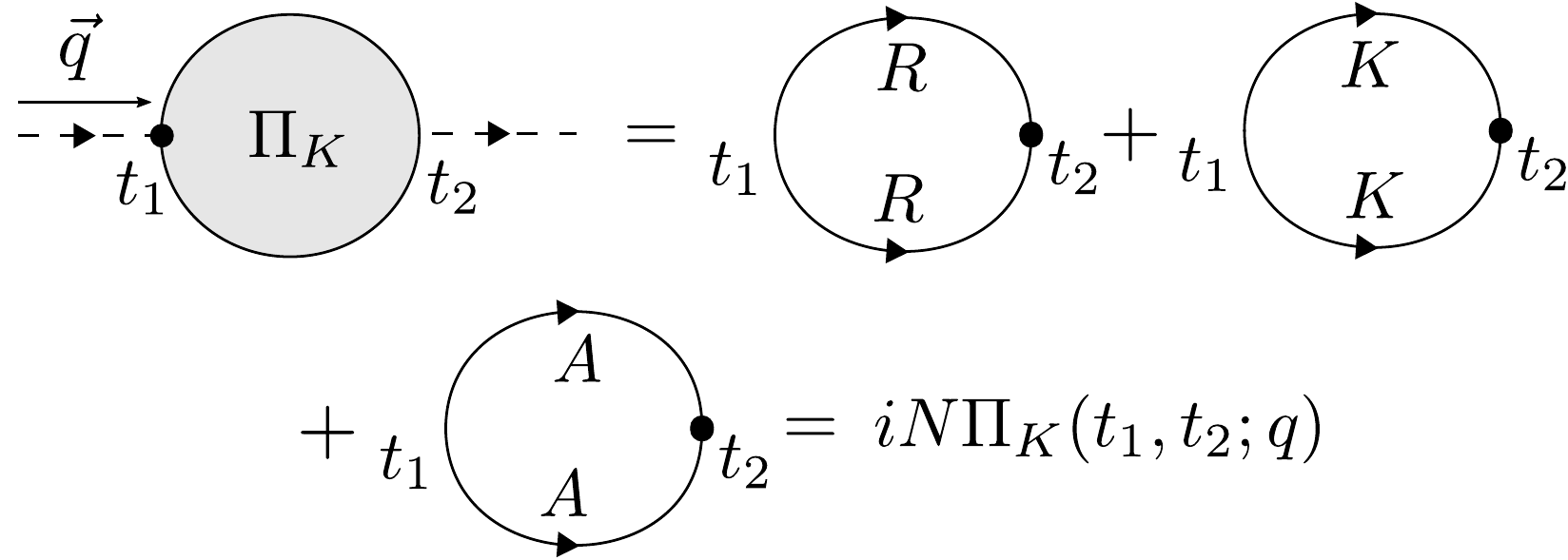}\\
\end{flalign*}
\caption{Definition of the $2PI$ generating functional: a) Schematic  of the $2PI$ functional $\Gamma'$. The Keldysh indices are suppressed for brevity. b) The retarded Cooperon. c) The Keldysh Cooperon. d) The retarded Cooper bubble. e) the Keldysh Cooper bubble. \label{fig:def2PI} }
\end{figure}

\begin{figure}
\begin{flalign*}
&{\raisebox{25pt}{}}\hspace{-0pt}\includegraphics[scale=.45]{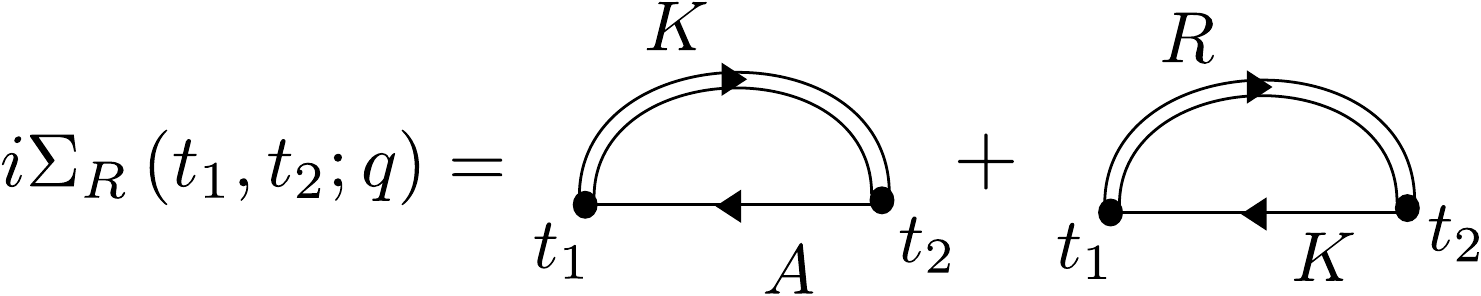}\\
\end{flalign*}
\caption{Retarded fermionic self-energy.\label{fig:SigmaR}}
\end{figure}

\begin{align}
G_R(1,2)&=-i\theta(t_1-t_2)\langle \{c(1), c^{\dagger}(2)\}\rangle,\nonumber\\
G_K(1,2)&=-i\langle [c(1), c^{\dagger}(2)]\rangle.
\end{align}
We here and generally suppress the spin and orbital indices, use numbers to indicate spacetime coordinates, and do not define the advanced function 
since it is the hermitian conjugate of the retarded part: $G_A(1,2) = G_R^*(2,1)$.

The  Green's function and interaction vertex correspond to the diagrams in Fig.~\ref{fig:FeynDefs}, while 
at ${\cal O}(1/N)$, the Keldysh functional $\Gamma'$ is the set of fermion loops shown in Fig.~\ref{fig:def2PI}. 
Note the effect of the $1/N$ expansion is to select the Cooper interaction channel which is the most singular channel
near the critical point.  

The Green's function is determined by the Dyson equation  
\begin{subequations}
\label{eq:EqOfMo}
\begin{equation}
{G_R}^{-1} = {g_R}^{-1}-\Sigma_R[G]\label{eq:SigToGR},
\end{equation}
where $g^{-1}$ is the non-interacting Green's function and $\Sigma_R$ is the retarded self-energy, determined self-consistently by the saddle point equation.
\begin{equation}
\Sigma_R[G] \equiv \delta\Gamma'/\delta G_A,
\end{equation}
At ${\cal O}(1/N)$, $\Sigma_R$ is (see Fig.~\ref{fig:SigmaR}),
\begin{equation}
\Sigma_R(1,2) =  \frac{i}{N}\left[D_K(1,2)G_A(2,1) + D_R(1,2)G_K(2,1)\right]\label{eq:defSigma}
\end{equation}
where $D$ is the Cooperon, Fig.~\ref{fig:def2PI}(b,c), defined by,
\begin{align}
D_R^{-1}&\equiv u^{-1}-\Pi_R.\label{eq:defDR}\\
D_K &\equiv D_R\circ \Pi_K \circ D_A. \label{eq:defDK}
\end{align}
Here $\Pi$ is the Cooper bubble, Fig.~\ref{fig:def2PI}(d,e), or equivalently the expectation,
\begin{align}
 i\Pi^K(q,t,t')&= \langle \{\Delta_q(t),\Delta^\dagger_q(t')\}\rangle, \label{eq:defPiK}\\
i\Pi^R(q,t,t')&= \theta(t-t')\langle \left[\Delta_q(t),\Delta^\dagger_q(t')\right]\rangle,\label{eq:defPiR}
\end{align}
\end{subequations}
evaluated to $\mathcal{O}\left(1/N\right)$.
The Cooperon can be understood as the correlator of an auxiliary or Hubbard-Stratonovich field $\phi$ conjugate to $\Delta$,
used to decouple the fermionic quartic interaction in the Cooper channel, as outlined in
Appendix~\ref{ap1}. In this language $D$ is defined by,

\begin{align}
D_R(1,2) &=-i\theta(t-t')\langle \left[\phi(1),\phi^*(2)\right]\rangle, \label{eq:defDRa}\\
D_K(1,2) &= -i\langle\left\{\phi(1),\phi^*(2)\right\}\rangle.\label{eq:defDK1}
\end{align}

We emphasize that Eqs.~(\ref{eq:EqOfMo}a-g), constitute a highly non-trivial set of coupled equations as $D$, $G$, and 
$\Sigma$ are defined self-consistently in terms of each other.  In the rest of this paper we solve these equations in the nonequilibrium system 
following the quench. In Sec.~\ref{sec:pert} we do this in a short time 
approximation, which gives the essential qualitative behavior. In Sec. \ref{sec:nonpert} we remove the short time limit and 
consider the general behavior.

\section{Perturbative Regime}\label{sec:pert}

We begin by evaluating Eqs.~\eqref{eq:EqOfMo} in the spirit of a short time approximation. We do this by replacing $G$ with it's initial,
noninteracting value $g$. The functions $D$ and $\Pi$ may then be straightforwardly obtained in terms of the dispersion $\epsilon_k$ and the initial 
occupation. At finite temperature $T$, and low frequency ($\omega/T\ll 1$) (see Appendix~\ref{ap2}) these can be estimated as, 
\begin{align}
\Pi_R(v_F|q| \ll T,\omega\ll T) &= \nu\left(a -ib\frac{\omega}{T} + a c^2 \frac{v_F^2q^2}{T^2}\right),\nonumber\\
i\Pi_K(v_F|q| \ll T,\omega \ll T)&= 4b\nu,\nonumber\\
\Rightarrow i\Pi_K(q=0,t,t')&\sim \delta(t-t'),\label{Pidef}
\end{align}
where $v_F$ is the Fermi velocity, $\nu$ the density of states and $a$, $b$ and $c$ system-dependent dimensionless constants. Then Eq.~\eqref{eq:defDR} reduces to
\begin{eqnarray}
&&\biggl[\partial_t + \gamma_q\biggr]D_R(q,t,t')=-Z\delta(t-t'),\nonumber\\
&&\gamma_q = T\left(l^2 q^2+r\right).
\label{eq:DREoM}
\end{eqnarray}
where we have $l = c v_F/T$ and $Z\sim T/\nu$, and $r$ is the distance from the critical point in units of $T$.

\begin{align}
D_R(q,t,t') &= -Z\theta(t-t') e^{-(t-t')\gamma_q},  \label{drg}
\end{align}
The $D$ are overdamped as a consequence of the fermionic bath. When $r >0$ ($r<0$) $D_R$ decays (grows) with time indicating that the system is 
in the disordered phase (unstable to the ordered phase). The critical point $r=0$ separates the two regimes.

While $D_{R,A}$ are time translation
invariant within the current approximation, $iD_K$ explicitly breaks time translation invariance. From Eqns.~\eqref{eq:defDK},~\eqref{Pidef}, and~\eqref{drg} 
it follows that
\begin{eqnarray}
iD_K(q,t,t')= Z\frac{T}{2\gamma_q}\biggl[e^{-\gamma_q|t-t'|}-e^{-\gamma_q(t+t')}\biggr].\label{dkg}
\end{eqnarray}
As expected of a non-equilibrium system this violates the fluctuation dissipation theorem. We may quantify this by introducing a function
\begin{equation}
F_K^0(x) = \frac{1-e^{-2x}}{2 x},\label{f0}
\end{equation}
Eq.~\eqref{dkg} may be written as
\begin{eqnarray}
&&iD_K(q,t,t')= -T\biggl[ D_R(q,t,t')t'F_K^0(\gamma_q t') \nonumber\\
&&+t F_K^0(\gamma_q t)D_A(q,t,t') \biggr].
\end{eqnarray}
The violation the FDT (at the initial temperature T) is given by the fact that,
\begin{equation}
2xF^{0}_K\!\left(x\right) - 1 \neq  0.
\end{equation}
Therefore this quantity measures the extent to which the Cooper pair fluctuations are out of equilibrium with the fermions.

At the critical point $r=0$, $iD_K\!\left(q,t,t\right)$ for $v_Fq \ll T$ can be written in the scaling form
\begin{equation}
iD_K\!\left(q,t,t\right)= Z T t F_K^0\!\left(T l^2q^2t\right); \quad v_Fq \ll T.\label{dkT1}
\end{equation}
This is a consequence of the fact that at the critical point the only length scale greater than $l$ is the one generated from $t$, $l\sqrt{T t}$.
 
\subsection{Fermion lifetime}
We now show that the growing fluctuations may be detected through the spectral properties of the fermions, in particular the lifetime. 
There is some subtlety with defining the lifetime, as the system is not translationally invariant and so the response functions may not be decomposed in frequency space. 
However, we may take advantage of the fact the rate of change of $D_K$ is $\gamma_q$, while the typical energy of a thermal fermion is 
$T\gg\gamma_q$.  Therefore it is reasonable to interpret the Wigner-Transform of the self energy,
\begin{equation}
\Sigma_R^{\rm WT}\left(k,\omega; t \right)
\equiv
\int d\tau e^{i\tau\omega}
\Sigma_R\left(k,t+\frac{\tau}{2}, t-\frac{\tau}{2}\right),
\end{equation}
as being the self energy at $\omega$ near time $t$, and the quantity,
\begin{equation}
\frac{1}{\tau\!\left(k,t\right)} \equiv \textrm{Im} \biggl[\Sigma_R^{\rm WT}\left(k,\omega=\varepsilon_k;t \right)\biggr], 
\end{equation}
as the fermion lifetime.
In any case, the correct observable will be determined by the particular experimental protocol.

We now proceed to evaluate $\tau^{-1}$ within the present approximation. The self energy is determined from equation~\eqref{eq:defSigma}
which in momentum space takes the form,
\begin{align}
\Sigma_R\left(k;t_1,t_2\right) &= 
i\!\int\!\! \frac{d^d q}{(2\pi)^d}\biggl[
	G_K\!\left(-k+q;t_2,t_1\right)
	D_R\!\left(q;t_1,t_2\right) 
	\nonumber\\
	& + 
	G_A\!\left(-k+q;t_2,t_1\right)
	D_K\!\left(q;t_1,t_2\right)
\biggr].
\end{align}
We may now make several simplifications. First as $D_K/ D_R \sim T/ \gamma_q$ we may neglect the first term. Second, 
as the evaluation of $D_K(q;t_1,t_2)$ has shown that it varies on the scale of $\gamma_q$ which is much less then $T$, 
it is sufficient to replace $D_K\!\left(q,t_1,t_2\right)$ with the equal time quantity $D_K(q;t,t)$, $t= (t_1+t_2)/2$ 
evaluated at the average value of the two time coordinates. Third, continuing within the perturbative approximation, 
$G_R$ may be replaced with it's non-interacting value. 

Therefore we obtain the equation, 
\begin{align}
\Sigma_R\!\left(k; t_1,t_2\right)&=\nonumber\\
 &\hspace{-30pt}i\theta\!\left(t_1-t_2\right)
\!\int\!\! \frac{d^d q}{(2\pi)^d} 
 e^{i\varepsilon_{q-k}(t_1-t_2)}
iD_K \!\left(q,t,t\right).
\end{align}

Finally, we Fourier transform with respect to the time difference $t_1-t_2$,
\begin{align}
\Sigma_R^{WT}\left(k,\omega;t \right)
&= -\!\int\!\!\frac{d^d q}{(2\pi)^d}
\frac {
	 iD_K\!\left(q,t,t\right)
}{
	\omega + \varepsilon_{q-k} +  i\delta
}.
\end{align}
Setting $\omega = \varepsilon_k$, and using Eq.~\eqref{dkT1}, we obtain,
\begin{align}
\Sigma^R\left(k,\omega = \varepsilon_k,t \right)
&= Z Tt \!\int\!\! \frac{d^d q}{(2\pi)^d}
\frac{ 
	F_K^0(\gamma_q t)}{\varepsilon_k + \varepsilon_{q-k}+i\delta}
\nonumber\\
&\approx Z Tt \!\int\!\! \frac{d^d q}{(2\pi)^d}
\frac{ 
	F_K^0(\gamma_q t)}{2\varepsilon_k + \vec{q}\cdot \vec{v}_k +i\delta}
\end{align}
Focusing on the region where $q l  <1$, we may set $\gamma_q = T l^2 q^2$. Going over to spherical coordinates this may be written as,
\begin{align}
\Sigma^R\left(k,\omega = \varepsilon_k,t \right)
&=Z T t\int_0^{l^{-1}}\!\!\!\!\!\frac{q^{d-1}d q }{(2\pi)^d}\bigg[
\nonumber\\
 &\hspace{10pt}\int\!\! d\hat{n}\frac{
F^0_K\!\left(tTl^2q^2\right)
	}{
	q \vec{v}_k\cdot \hat{n} + 2\varepsilon_k +i\delta}\bigg],
\end{align}
where $\int d\hat{n}$ indicates the integral over the $d$-dimensional sphere.  Now introducing new coordinates $y = q l \sqrt{Tt}$, produces 
\begin{align}
\Sigma^R\left(k,\omega = \varepsilon_k,t \right)
&=(Tt)^\frac{3-d}{2}\frac{Z c}{Tl^d}
\nonumber\\
&\hspace{-30pt}\times\!\! \int_0^{\sqrt{Tt}}\!\!\frac{y^{d-1} d y}{(2\pi)^d}\!\! \int\! d\hat{n}\frac{
F^0_K\!\left(y^2\right)
	}{
	 y \cos\theta' + 2c\varepsilon_k\sqrt{t/T} +i\delta}\nonumber\\
&=(Tt)^\frac{3-d}{2}\frac{Z c}{Tl^d}S^0_d\left(2 c \varepsilon_k\sqrt{t/T}\right),\end{align}
\begin{align}
S^0_d\!\left(x\right) &\equiv \int_0^{\sqrt{Tt}}\!\frac{y^{d-1} d y}{(2\pi)^d}\!\! \int d\hat{n}\frac{
F^0_K\!\left(y^2\right)
	}{
	 y \cos \theta' +x +i\delta},
\label{eq:defS0}
\end{align}
where $\theta'$ is the angle between $\vec{v}_k$ and $\hat{n}$. Therefore the lifetime is given by
\begin{align}
\tau^{-1}\!\left(k,t\right) = \left(Tt\right)^\frac{3-d}{2} \frac{Z c}{T l^d} \text{Im} S^0_d\left(2 c \varepsilon_k\sqrt{t/T}\right).
\label{eq:reltauS}
\end{align}
%Straightforward manipulation gives that
%\begin{equation}
%\text{Im} S^0_d(x)=\frac{1}{2}\int \frac{d^{d-1}y}{(2\pi)^{d-1}} F_K^0\left(\left[y^2 + x\right] \right).
%\end{equation} 

\begin{figure}
\includegraphics[width = .95\columnwidth]{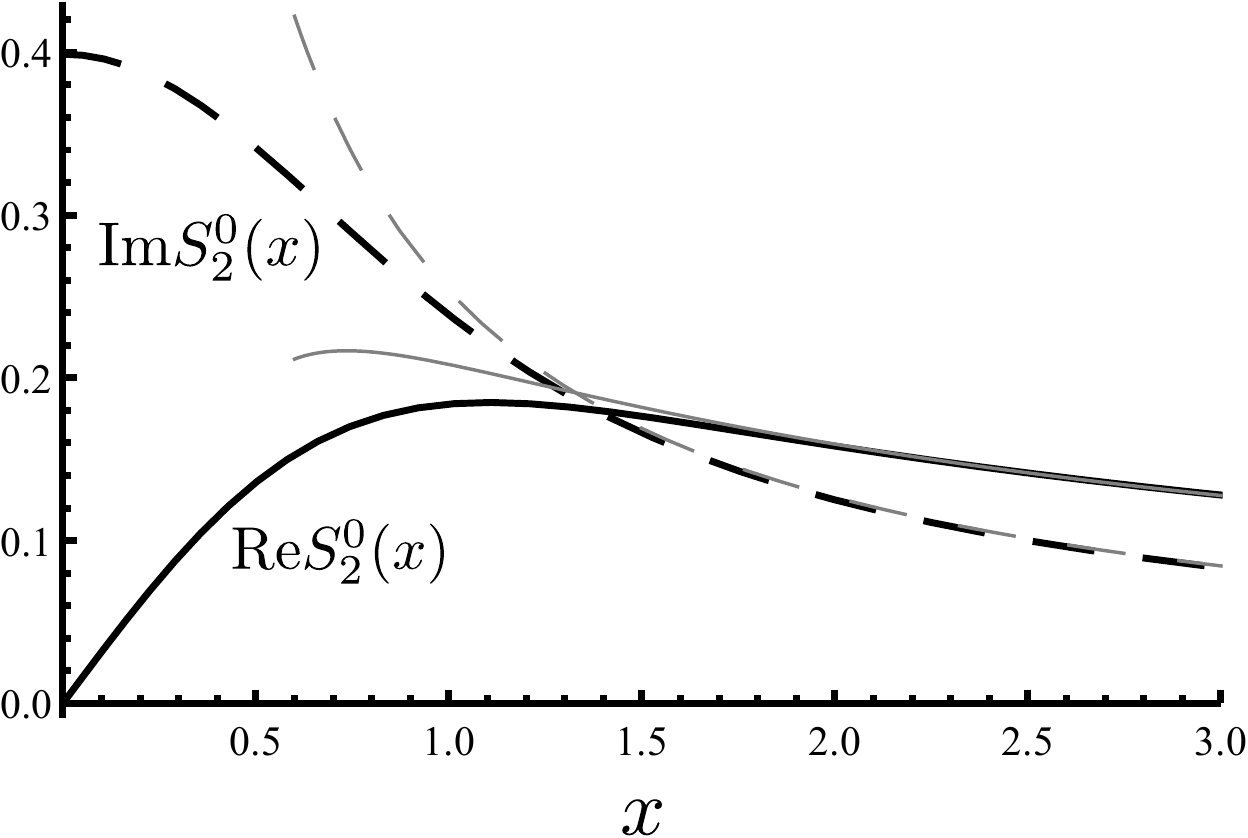}
\caption{The real and imaginary parts of the scaling function $S_2^0$ defined in Eq.~(\ref{eq:defS0}). 
The lighter lines show the asymptotic predictions, Eq.~(\ref{eq:defS02asymp}) and Eq.~\eqref{eq:RdefS02asymp}. The asymptotic forms are extremely accurate for arguments greater than one. 
}
\label{fig:S2}
\end{figure}

We now evaluate this function, beginning in $d=2$.  
	Recalling that $F^0_K(y)\rightarrow y^{-1}$ as $y\rightarrow\infty$ and $F^0_K(0) = 1$, we see that in $d =2$ the integral 
is convergent and so we may neglect the condition that $q \ll l^{-1}$. Therefore $S^0_2(x)$ is not sensitive to how the 
integral is cutoff at $q \approx l^{-1}$, and is plotted in Fig.~\ref{fig:S2}. The asymptotics may be extracted from the asymptotics of $F^0_K$,
\begin{align}
\textrm{Im} S^0_2(x) &\sim \text{const};\qquad &x\ll 1 \nonumber\\
		 &\sim \frac{1}{x};\qquad &x \gg 1.  
\label{eq:defS02asymp}
\end{align}

For completeness we give the asymptotic forms of {\rm Re}$S_2^0$ which take the form. 
\begin{align}
\text{Re} S_2^0(x) &\propto x; & x\ll 1 \nonumber \\
&= \frac{1}{2x}\log\left(x/\sqrt{T t}\right); & x\gg 1. \label{eq:RdefS02asymp}
\end{align}

As a result, for $d=2$,
\begin{align}
\tau^{-1}\!\left(k,t \right) &\sim \sqrt{t};& t\varepsilon^2_k/T\ll 1 \nonumber\\
&\sim \frac{1}{\varepsilon_k};& t\varepsilon^2_k/T \gg 1 .\nonumber\\
\end{align}
The full curve is plotted in Fig.~\ref{fig:S2}.

In $d = 3$ the integral is logarithmically divergent. The upper limit of the integral over $y$ is $\sqrt{t T}$ whereas the divergence as $y
\rightarrow 0$ is cutoff by the greater of $x$ and one. Thus, 
$S^0_3(x) \sim \log\left[ \sqrt{Tt}/{\rm max}(1,x)\right]$, and
\begin{align}
\textrm{Im} S^0_3(x) &\sim \log(T t); \qquad &x\ll 1 \nonumber\\
		 &\sim \log\biggl(\frac{T t}{x^2}\biggr);\qquad &x \gg 1.
\label{eq:S03asymp}  
\end{align}

For completeness we give asymptotic limits of {\rm Re}$S_3^0$ which takes the form.  
\begin{align}
\text{Re} S_3^0\left(x\right)& \sim x; &x \ll 1 \nonumber \\
&\sim \text{const} & x \gg 1. \label{eq:RS03asymp}
\end{align}

\begin{figure}
\includegraphics[width = .95\columnwidth]{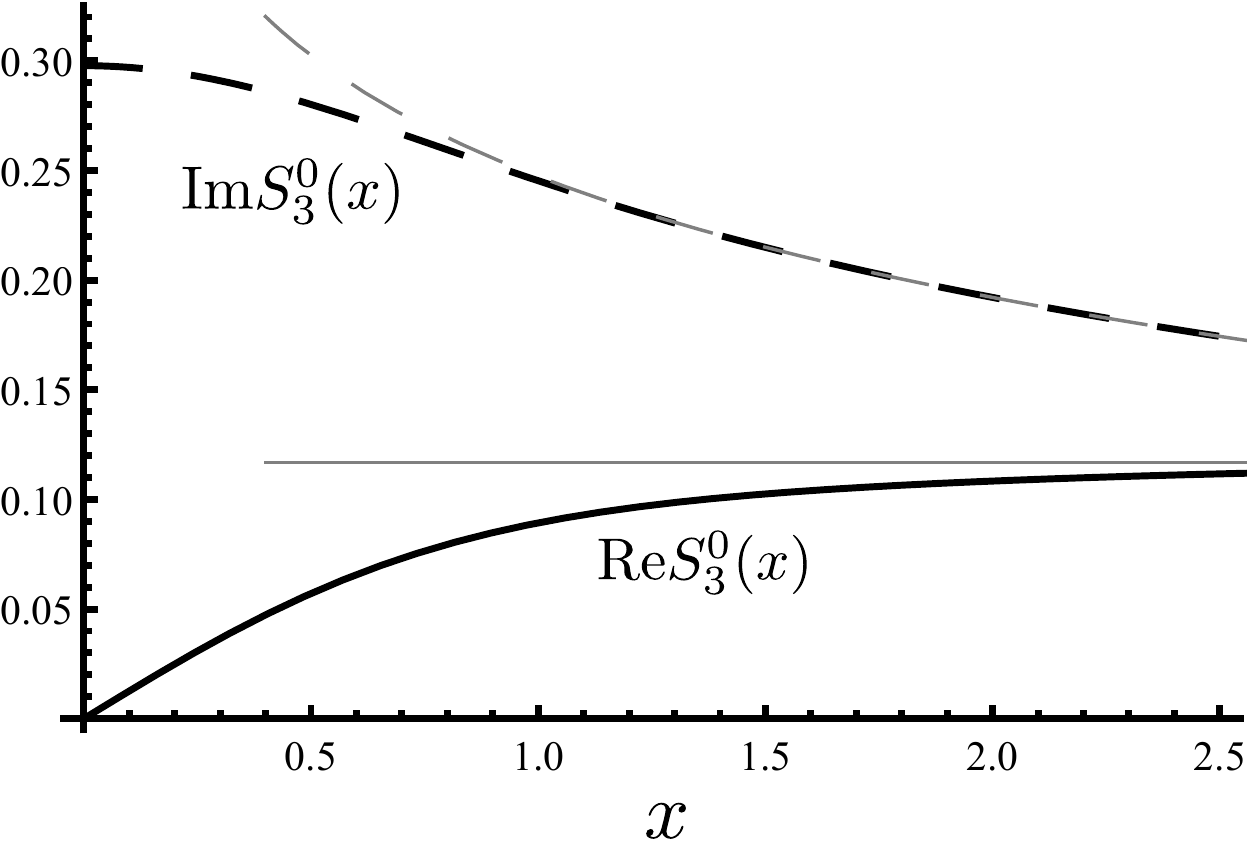}
\caption{The real and imaginary parts of the scaling function $S^0_3$ defined in Eq.~(\ref{eq:defS0}) for purely real arguments $x$. 
The lighter lines show the asymptotic predictions, Eq.~(\ref{eq:S03asymp}) and Eq.~\eqref{eq:RS03asymp}. The asymptotic forms are extremely accurate for arguments greater than one. Note that $\text{Im}S^0_3$ is only defined up to an additive constant, see discussion after Eq.~(\ref{eq:S03asymp})
}
\label{fig:S3}
\end{figure}

Therefore for $d=3$, we have the dependence,
\begin{align}
\tau^{-1}\!\left(k,t \right) &\sim \log\left(T t\right);& t\varepsilon^2_k/T\ll 1 \nonumber\\
&\sim - \log\left(\frac{\varepsilon_k}{T}\right);& t\varepsilon^2_k/T\gg 1.\nonumber\\
\end{align}
The full behavior of the function $S^0_3$ is plotted in Fig.~\ref{fig:S3}. Note that as this function is logarithmically  
dependent on the cutoff of the integral, a change in the precise form of the cutoff enforcing $q \ell \ll 1$ will shift the final result by an additive constant. This ambiguity would be fixed by comparing the asymptotic behavior of $\tau(\varepsilon)$, $\varepsilon\rightarrow \infty$ with a microscopic calculation of the lifetime at high energies.

The self-energy at $\varepsilon = 0$ diverges as $t\rightarrow\infty$. Therefore at sufficiently long time the assumption that $G$ may be substituted with it's non-interacting value is not valid. In the next section we lift this assumption.

\section{Non-Perturbative Regime}\label{sec:nonpert}

We note that the $2PI$ formalism is not a short-time expansion, and that the Eqns.~\eqref{eq:EqOfMo} are valid at all times. There are two shortcomings that must be remedied. 

First, we have seen that the $D$ is a slow function of time, and that $\Sigma$ depends essentially on the equal time value $D_K(t,t)$. The estimation in Sec.~\ref{sec:pert} 
essentially estimates $D_K(t,t)$ by linear response. However as $D$ increases with time, it will eventually grow large enough to violate the linear response assumption. 
We remedy this by employing the theory of critical quenches in this section.

Second, we have that $G$ and $\Sigma$ must be self-consistent. We therefore solve the self-consistent version of Eq.~\eqref{eq:EqOfMo}. 
The results of this analysis are shown in Fig.~\ref{fig:summary}.

\subsection{Propagator for interacting Cooperons}\label{critquench}

The equation of motion for $D$ given in Eq.~(\ref{eq:DREoM}), is equivalent to linear response. This is because the non-linear behavior comes from the 
dependence of $G$ on $\Sigma$, which in turns depends on $D$. Therefore once the perturbative approximation for $G$ fails it becomes necessary to consider 
the non-linear evolution of $D$. 

Since we are only concerned with the long wavelength behavior of $D$ in the vicinity of the dynamical critical point, we do not need to solve for the full behavior of $D$. Instead we need only to identify the appropriate dynamical universality class. As the fluctuations of the bosonic field $\Delta$ are not conserved and are overdamped by the fermionic bath, the system belongs to the dynamical model-A transition~\cite{HH77}. The standard manipulation mapping the original Hamiltonian to this model are relegated 
to Appendices~\ref{ap1},~\ref{ap2}.

We quote the results for the transient dynamics of thermal aging in model A 
already discussed elsewhere~\cite{Janssen1988,Gambassi05},
and re-derived in Appendix~\ref{ap3}. 

In $d=2$ there is no true dynamical critical point, as in equilibrium. Therefore our results are only valid in the perturbative regime in $d=2$. 
The behavior at intermediate to long time is expected to be described by a crossover to Kosterlitz-Thouless physics, where the 
amplitude of the Cooper fluctuations saturate but there are long range phase fluctuations. However, such a calculation is beyond the scope of this paper.

In $d=3$, the dynamics is characterized by three exponents $z,\eta,\theta$. Of these $z,\eta$ are already
familiar in equilibrium $\phi^4$ theory and are the dynamical critical exponent and the
scaling dimension of $\phi$ respectively. 

The exponent $\theta$ is a non-equilibrium exponent 
known as the initial slip exponent~\cite{Janssen1988}. It is responsible for non-trivial
aging dynamics, and is interpreted as the scaling
dimension of a source field applied at short times after the quench. This is because such a source field will
induce an initial order-parameter $M_0=\langle \phi_c\rangle$, to  
grow with time at short times after the quench 
as~\cite{Janssen1988} $M_0 \sim t^{\theta}$ even though the quench is still within the disordered phase.  
At long times, eventually the order-parameter will decay to zero. 
For the present calculation, the short time behavior is not directly relevant as we are interested in the regime when $|t-t'| \ll t$. 
However the short-time exponent still affects the qualitative behavior of $D_K$.

The values of the Model A exponents $z$, $\eta$ and $\theta$ may be calculated using standard methods such as epsilon-expansions or large-N,
where $N$ now controls the components of the bosonic field. We adopt the latter approach, and emphasize that this component $N$ 
is not the same as the fermion orbital index used to justify the form of $\Gamma'$. 
The derivation of the exponent using large-N for the bosonic theory is equivalent to a Hartree-Fock approximation for model A
(see App.~\ref{ap3}) giving $z=2$, $\eta=0$  and $\theta = \epsilon/4, \epsilon=4-d$.  
Other approximations will change the precise value of the exponent, but not
the overall scaling form.

The results for the Cooperon dynamics from model A (see App.~\ref{ap3}) are as follows. When $t'$ becomes comparable to $t$, we expect,
\begin{align}
iD_K(q,t,t') &= T t' e^{-q^2(t-t')}\left(t/t'\right)^{\theta}F_K(2q^2t'); \quad q l \ll 1,\nonumber\\
F_K(x)& \equiv \int_0^1 dy e^{-xy}(1-y)^{-2\theta}.\label{FKdefm}
\end{align}
In particular for equal times,
\begin{eqnarray}
&&i D_K(q,t,t)\propto t F_K(2 q^2t).
\end{eqnarray}

The above form for the boson density $iD_K(q,t,t)$ is the same as that in Sec.~\ref{sec:pert} if one replaces 
the scaling function $F_K^0$ by $F_K$. The function $F_K$ has the asymptotic limits

\begin{align}
F_K(x) &= (1-2\theta)^{-1}; & x = 0\nonumber\\ 
	   &\sim x^{-1} + 2\theta x^{-2};& x\gg 1.
\end{align}
Note the leading asymptotic behavior of $F_K(x)$ as $x\rightarrow \infty$ is the same as that of $F_K^0(x)$.

We define a scaling function $S_d$, $d>2$ as the analogue of Eq.~(\ref{eq:defS0}) for interacting bosons,
\begin{equation}
S_d\!\left(x\right) \equiv \int_0^{\sqrt{Tt}}\!\frac{d y y^{d-1}}{(2\pi)^d}\!\! \int d\hat{n}\frac{
F_K\!\left(y^2\right)
	}{
	 y \cos \theta' +x +i\delta}.
\end{equation} This replacement makes only a small 
quantitative change to the final result in $d=3$. As the leading behavior at $F_K$ is unchanged, 
the derivation of the asymptotics Sec. ~\ref{sec:pert} may be followed precisely, leading to
\begin{align}
{\rm Im}S_3(x) &\sim \text{const};    & x\ll 1\nonumber\\
		&\sim \log\left( Tt/x^2\right)  + \text{const}; &x \gg 1.
\label{eq:newSAsymp}
\end{align}  
The only difference in the asymptotic behavior between ${\rm Im}S_3^0$ and ${\rm Im}S_3$ might be in the constants. However $\tau^{-1}$ 
only depends on ${\rm Im}S_3\left( \beta \right)$ where $\beta$ is a material dependent parameter, see  Eq.~(\ref{eq:reltauS}). 
Further as ${\rm Im}S_3$ has a logarithmic dependence on the cutoff, changing the (material-dependent) cutoff 
shifts ${\rm Im}S_3(x) \rightarrow {\rm Im}S_3(x) + \gamma$. Thus the constants in Eq.~(\ref{eq:newSAsymp}) may be absorbed into constant $\beta, \gamma$. 

\begin{figure}
\includegraphics[width = .95\columnwidth]{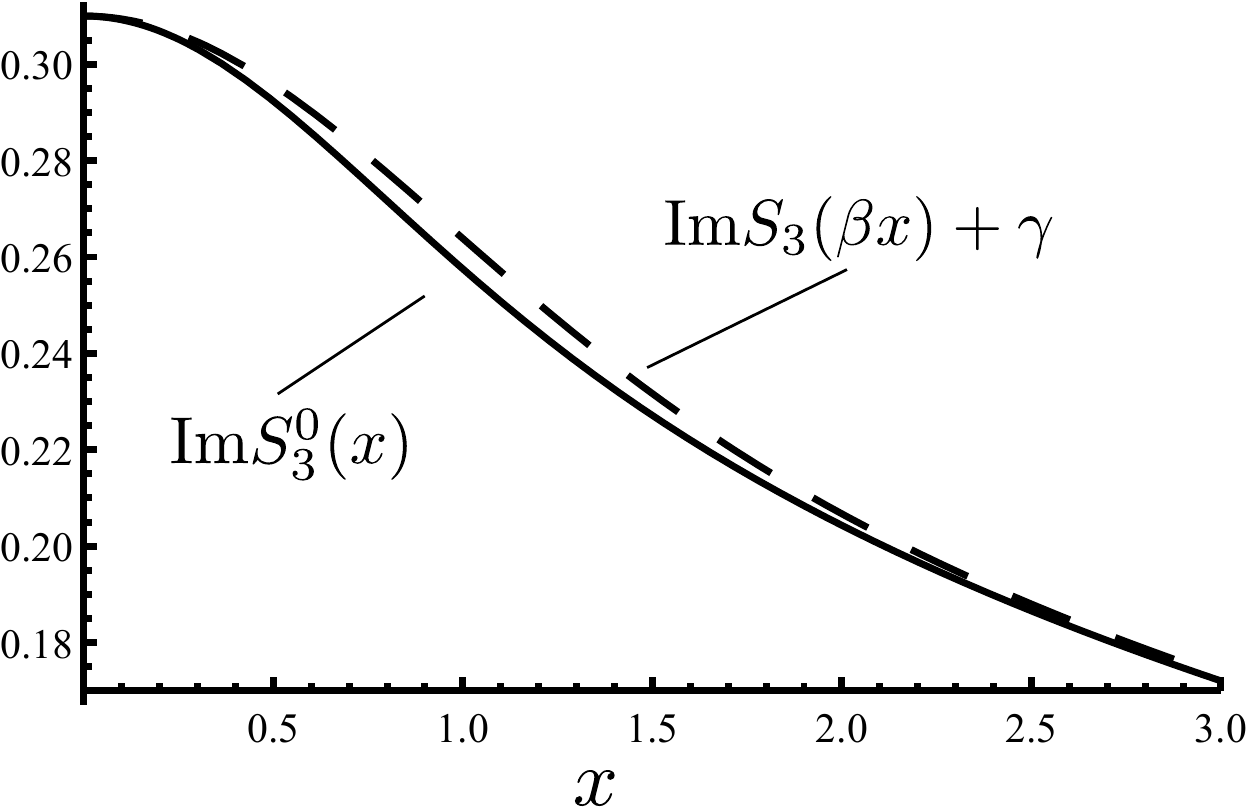}
\caption{Plot of $\text{Im}S_3^0$ (full line) and $\text{Im} S_3$ (dashed line). The constants $\beta$ and $\gamma$ are cutoff dependent constants chosen so that the functions agree at $x = 0$ and as $x\rightarrow\infty$.}
\label{fig:compIntFree}
\end{figure}

Once these are fixed, ${\rm Im}S_3(x)$ and ${\rm Im}S_3^0(x)$ both crossover smoothly between the same asymptotics and therefore it is reasonable that they are qualitatively similar, see Fig.~\ref{fig:compIntFree}. As the final result does not appear significantly sensitive to the critical exponents we will not attempt to estimate the value of these exponents more accurately.

\subsection{Self-consistent solution}
Although the intermediate regime is only controlled for the case of $d=3$, we include the self-consistent equation in both $d=2,3$ 
for completeness. Having considered the behavior of $D$ we may now directly solve the self consistent equation for the self-energy, Eq.~\eqref{eq:EqOfMo}.  
In Wigner coordinates, it is given by,
\begin{equation}
\Sigma^R(k,\omega,t) = i\int \frac{d^d q}{(2\pi)^d} \frac{D_K\left(q,t,t\right)}{\omega + \epsilon_{k+q} + \Sigma^R(k+q,\omega,t)}.
\end{equation}

 Expanding the denominator in $q$ to first order, and  assuming that the variation in $\Sigma^R$ with $q$ is negligible, gives
\begin{equation}
\Sigma^R(k,\omega,t) = i\int \frac{d^dq}{(2\pi)^d} \frac{D_K\left(q,t,t\right)}{\omega + \epsilon_k + \Sigma^R(k,\omega,t) + \vec{v}_k\cdot \vec{q} }.
\end{equation}

Now the scaling for $D_K$ is 
\begin{equation}
i D_K(q,t,t) =  Z T t F_K\biggl(\frac{2v^2 q^2 t}{T}\biggr).
\end{equation}
To take advantage of this we rescale units by defining
\begin{align}
y &\equiv v |q| \sqrt{2 t/ T} \\
z_0 &\equiv \left(\omega + \varepsilon_k + i\delta\right)\sqrt{t/T}\\
z &\equiv \left(\omega + \varepsilon_k + \Sigma^R(k,\omega,t)\right)\sqrt{t/T}\\
\alpha &\equiv \frac{Z}{T^2 l^d}. 
\end{align}
Giving
\begin{equation}
z = z_0 + \alpha (T t)^{\frac{4-d}{2}}\int_0^{y_m} \frac{d y y^{d-1}}{(2\pi)^d} \int d\hat{n} \frac{F_K(y^2)}{y\hat{k}\cdot\hat{n} + z}.
\label{eq:dmlessSC}
\end{equation}
The integral over $\hat{n}$ is an integral over unit vectors in $\mathbb{R}^d$. The condition that $v|q| \ll T$ is imposed by cutting off the integral at $y_m \sim \sqrt{t T}$. 
So, it goes to infinity as $t \rightarrow \infty$. This gives a self-consistent equation for $z$.  The function $F_K(w)$ goes to a constant as 
$w\rightarrow 0$, decays like $1/w$ as $w\rightarrow \infty$. We introduce the function $S_d$ depending on dimension $d$ so that we can write,
\begin{equation}
z = z_0 + \alpha (Tt)^{\frac{4-d}{2}}S_d(z).
\end{equation}
Since we need to solve the integral self-consistently we must understand how $S_d(z)$ behaves for $z$ in the upper half of the complex plane.  
This is greatly simplified since $S_d$ is analytic as a function of $z$ in the upper half complex plane, as the only singularity can come from the pole 
$y \hat{k}\cdot\hat{n}= -z$. The estimate of $S_d$ depends on the dimensions. 

\subsubsection{ $d=3$ }
In $d=3$, Eq.~(\ref{eq:dmlessSC}) leads to 
\begin{align}
S_3(z) &= \frac{1}{4\pi^2}\int_0^{y_m} dy y^2 F_K(y^2) \int_{-1}^{1} \frac{d\cos\theta'}{y\cos\theta' + z}\\
&= \frac{1}{4\pi^2} \int_0^{y_m} dy y F_K(y^2)\log\left(\frac{z+y}{z-y}\right).\
\label{eq:defS3}
\end{align}

Recalling the position of the branch cut as $z\rightarrow i\delta$ we obtain 
\begin{equation}
\log\left(\frac{i\delta + y}{i\delta - y}\right) = \pi i,
\end{equation}
so that as $|z| \rightarrow 0$ we get the leading behavior as $y_m\rightarrow \infty$,
\begin{align}
S_3(z) &=  \frac{i}{4\pi} \int_0^{y_m} dy y F_K(y^2)\\
	&\sim \frac{i}{4\pi} \log(y_m/\zeta),
\end{align}
where $\zeta$ is an order one constant.
The integral does not converge as $y_m$ goes to $\infty$. Therefore, $S_3$ does not depend only on the variable $z$ but also on $y_m$ and therefore exactly how the 
integral is cutoff at $q\approx q_m$. 
In particular by shifting $y_m$ to a new value $y'_m$ changes $S_3 \rightarrow S_3 + i\log( y_m/y'_m)/4\pi$. Therefore the imaginary part 
of $S_3$ is ambiguous up to an overall additive constant. 

To understand the large $z$ behavior, we split the integral into the regions $y \ll \zeta$ and $ y\gg \zeta$, where $\zeta$ is some constant of order one. The small $y$ limit is
\begin{align}
\int_0^\zeta dy y F_K(y^2)& \log\left(\frac{z+y}{z-y}\right)\nonumber \\ &\sim \int_0^\zeta dy y F_K(y^2)\left[ 1 + \frac{2y}{z} +\cdots\right]\\
	&\sim \textrm{const}.
\end{align}

And the large $y$ limit is
\begin{align}
\int_\zeta^{y_m} &dy y F_K(y^2)  \log\left(\frac{z+y}{z-y}\right) \approx \int_\zeta^{y_m} \frac{dy}{y}  \log\left(\frac{z+y}{z-y}\right)\\
&=\pi i\log\frac{y_m}{\zeta} + \int_\zeta^{y_m} \frac{dy}{y}  \left[\log\left(\frac{z+y}{z-y}\right) - \pi i\right]\\
&\approx \pi i\log\frac{y_m}{\zeta} + \int_\zeta^{\infty} \frac{dy}{y}  \left[\log\left(\frac{z+y}{z-y}\right) - \pi i\right]\\
&\approx \pi i \log\frac{y_m}{\zeta} + \int_{\zeta/z}^{\infty} \frac{du}{u}  \left[\log\left(\frac{1+u}{1-u}\right) - \pi i\right]
\end{align}

As $z \rightarrow \infty$ this diverges logarithmically around $u = 0$, therefore the integral is approximately $-\pi i\log\left(\zeta/z\right)$. Collecting the results we have that,

\begin{align}
{\rm Im}S_3(z)  &= \frac{i}{4\pi} \log\left(\frac{y_m}{\zeta}\right) +\cdots; \qquad &z\rightarrow 0&\\
		&= \frac{i}{4\pi} \log\left(\frac{y_m}{z}\right); \qquad &z\rightarrow\infty
\label{eq:S3asymp}
\end{align}  
The results are summarized in Fig.~\ref{fig:S3}. Note the the substitution of $F_K$ for $F_K^0$, makes minimal difference in the calculation of $S_3$, 
see Fig.~\ref{fig:compIntFree}. 

Returning to the self consistent equation
\begin{align}
z = z_0 + \alpha (Tt)^{\frac{4-d}{2}} S_3(z)
\label{eq:SCz}
\end{align}
If we assume that $z \approx z_0$, we obtain
\begin{align}
z = z_0 + \alpha (T t)^{\frac{4-d}{2}} S_3(z_0)
\end{align}
Plugging this back into the self consistent equation
\begin{align}
z &= z_0 + \alpha (T t)^{\frac{4-d}{2}} S_3\left[
z_0 + \alpha (Tt)^{\frac{4-d}{2}} S_3(z_0)
\right]\nonumber\\
&\approx %z_0 + \alpha (T t)^{\frac{4-d}{2}} S_3(z_0)\nonumber \\
%&\quad +   \left(\alpha (T t)^{\frac{4-d}{2}} S_3(z_0)\right)\left[ \alpha (Tt)^{\frac{4-d}{2}} S'_3(z_0)\right]\nonumber\\
%&=  
z_0 + \alpha (T t)^{\frac{4-d}{2}} S_3(z_0) \left(1+ \alpha (T t)^{\frac{4-d}{2}} S'_3(z_0)\right).
\end{align}
This implies the condition for validity of the perturbative solution is 
\begin{equation}
1\gg \alpha (T t)^{1/2}S'_3(z_0) \sim \alpha (T t)^{1/2}/z_0 
\end{equation}

Substituting $z_0 = 2\varepsilon\sqrt{t/T}$, we see this condition is equivalent to,
\begin{equation}
\varepsilon \gg \alpha T.
\end{equation}
Therefore the short time dynamics is sufficient to explain the behavior of the tails of the distribution, which is reasonable as these saturate at short times.

Let us look for the self-consistent solution at $z_0 = 0 $ and $\alpha (T t)^{1/2} \gg 1$. Assuming $z\gg 1$ we get

\begin{equation}
z = \frac{\alpha}{4\pi} (T t)^{1/2} i \log(y_m/z).
\end{equation}
Bearing in mind the $y_m \propto \sqrt{t}$ we see that the above has a solution with $z \propto t^{1/2}$. 
Therefore the $\Sigma_R(0,0,t)$ saturates at a constant at long times, given by the equation
\begin{equation}
\frac{\Sigma^R(0,0,\infty)}{T} = -i \frac{\alpha}{4\pi} \log \frac{T}{\Sigma_R(0,0,\infty)}.
\end{equation}

To summarize, for $\varepsilon_k/T \gg \alpha$, the behavior is the same as in Sec. \ref{sec:pert}, with saturation at $\log \varepsilon_k / T$.  
For smaller energies the logarithmic growth given earlier saturates at   $T t \sim \alpha^{-2}$. 
The general behavior is shown in Fig.~\ref{fig:summary}. 

\begin{figure}
\includegraphics[width = .95 \columnwidth]{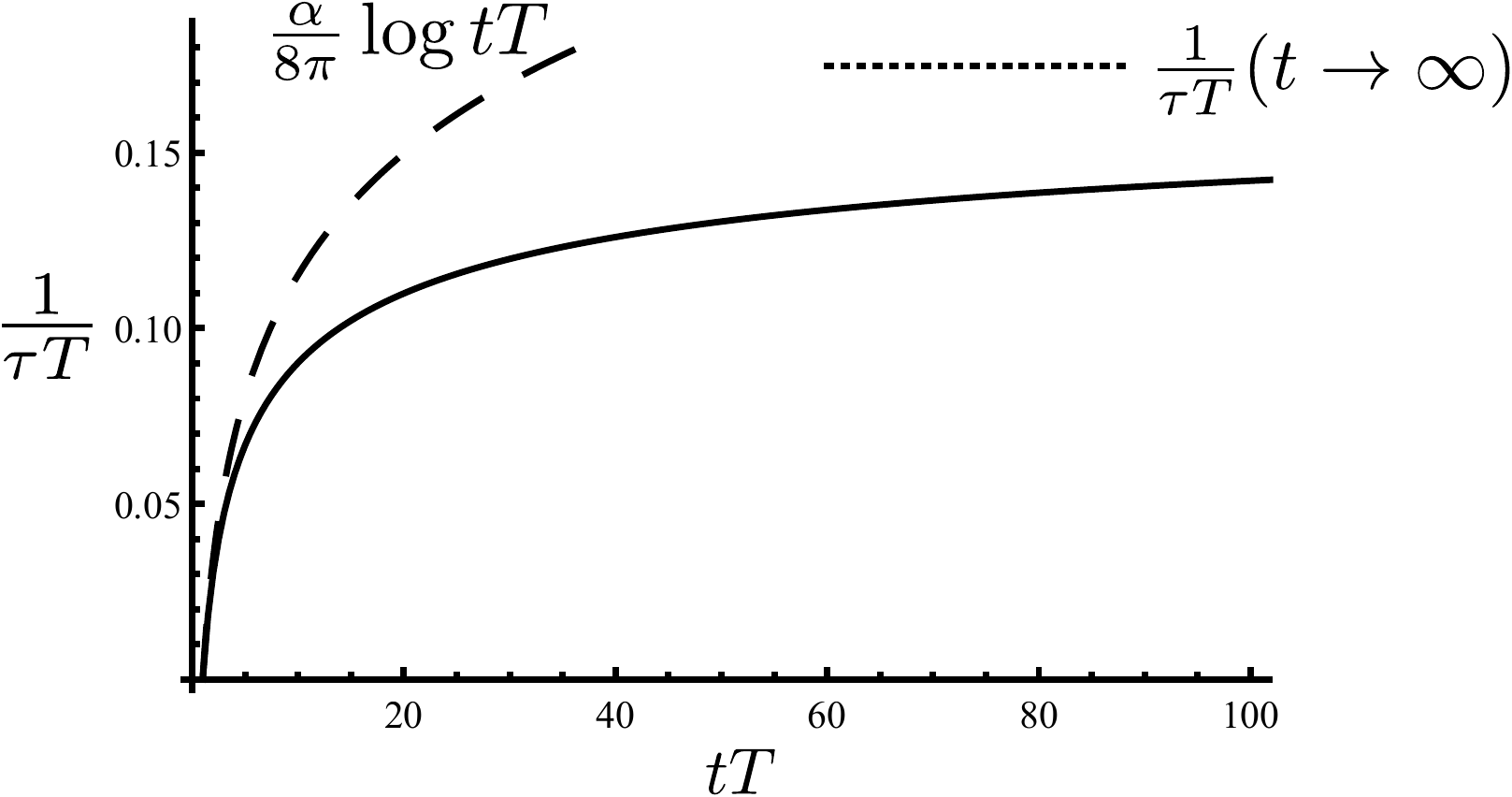}
\caption{The full line gives the growth of $\tau^{-1}(\varepsilon = 0)$ as a function of time since the quench. The curve is calculated following the discussion around Eq.~(\ref{eq:S3approx}) using the parameter $\alpha/4\pi = 0.1$, $\zeta = 1$ and $y_m = \sqrt{T t}$. The dashed and dotted lines show the asymptotic predictions for the short and long time regions respectively. The function only converges logarithmically to it's long time limit as $t\rightarrow\infty$, therefore it is still quite far from its asymptote on 
this scale.}
\label{fig:LongTimeLifeTime}
\end{figure}
 The approximate behavior of $\tau^{-1}$ at $\varepsilon_k =0$ is shown in  Fig.~\ref{fig:LongTimeLifeTime}.  Unfortunately calculating $S_3(z)$ over the upper half plane and solving Eq.~\eqref{eq:SCz} is numerically intensive. Instead we approximate
\begin{equation}
 S_3(z)\sim \frac{i}{4\pi}\log \frac{y_m}{\zeta + i z},
 \label{eq:S3approx}
\end{equation}
which renders Eq.~\eqref{eq:SCz} analytically tractable.
As this approximation has the same asymptotic limits as $S_3$ it should be sufficient for reproducing the qualitative shape of $\tau^{-1}$. 

\subsubsection{ $d=2$ }
We now analyze the self consistent equation in $d=2$. 

\begin{align}
S_2(z) &= \int_0^\infty dy \frac{y F_K(y^2)}{(2\pi)^2} \int_0^{2\pi} \frac{d\theta'}{y\cos\theta' + z}\\
&= \frac{1}{2\pi} \int_0^\infty dy \frac{y F_K(y^2)}{\sqrt{z^2- y^2}}\
\label{eq:defS2}
\end{align}
We estimate this integral as follows. First as $z \rightarrow 0+ i\delta $ this goes to $i$ for some order one constant. 
Note the sign is determined by the branch cut and should be consistent with causality. On the other hand if $z\gg 1$ 
we split the integral at some $\zeta$ of order one:
\begin{align}
\int_0^\zeta dy \frac{y F_K(y^2)}{\sqrt{z^2- y^2}} &\approx \frac{1}{z}\int_0^\zeta dy y F_K(y^2)\left[1 + \frac{y^2}{2z^2} + \cdots\right]\\ 
&\propto \frac{1}{z}, 
\end{align}
and the other half
\begin{align}
\int_\zeta^\infty dy \frac{y F_K(y^2)}{\sqrt{z^2- y^2}} &\approx \int_\zeta^\infty \frac{dy}{y\sqrt{z^2- y^2}}\\ 
&= \frac{\log (i \zeta)}{z}-\frac{\log \left(z +  \sqrt{z^2-\zeta^2}\right)}{z}.
\end{align}

As $|z|\rightarrow \infty$ this is $\sim \log(2z/\zeta)/z$, which dominates the small $y$ contribution, and it's effect is to renormalize the 
order one cutoff $\zeta$. So we may summarize the behavior as 

\begin{align}
S_2(z)  &= i\cdot \text{const}; \qquad &z\rightarrow 0&\\
		&= -\frac{1}{2\pi z}\log(\frac{z}{i \zeta}); \qquad &z\rightarrow\infty& 
		\label{eq:S2asymp}	
\end{align} 
 
The real and imaginary parts of $S_2^0$ which is similar to $S_2$, are plotted in Fig.~\ref{fig:S2}.
We deal with the self consistent equation essentially as in $d=3$. The perturbative condition holds at large $z_0$,
\begin{equation}
 |\alpha T t S'_2(z_0)|\ll 1 .
\end{equation}
For $z_0 = 0$ this condition is always violated at the time scale $1/\alpha$. However, if we take $z_0 = 2\varepsilon \sqrt{t/T} \gg 1$, then using the asymptotics we estimate that 
\begin{equation}
 \alpha T t S'_2(z_0)\sim \frac{\alpha T t \log(z_0)}{z_0^2} = \frac{\alpha T t \log\left(2\varepsilon\sqrt{t/T}\right)}{4\varepsilon^2 t/T}.
\end{equation}

The perturbative condition is only violated at an exponentially long time $t \propto \exp( \varepsilon^2/(\alpha T^2)$ ).

We now seek a self-consistent solution when $\alpha T t \gg 1$, but $z_0$ is small.  We use the large $z$ asymptotics.
\begin{equation}
z \approx z_0 - \frac{2\pi \alpha T t}{z} \log(z).
\end{equation}
Solving the quadratic equation treating $\log(z)$ as a constant we get,
\begin{equation}
z= \frac{z_0}{2}\left(1 + \sqrt{1-\frac{8\pi  \alpha t }{z_0^2}\log z}\right).
\end{equation}
The choice of branch comes from matching the behavior as $\alpha \rightarrow 0$. In the regime of interest where $t\gg 1$, 
we can to good accuracy simply replace the $\log z$ on the RHS with $\log (\pi \alpha T t)$. 

Taking the $t \gg 1 $ limit we obtain
\begin{equation}
z = \left[-\pi\alpha t T\log\left(- \pi\alpha t T\right)\right]^{1/2} .
\end{equation}

We see that $z\gg 1$ so the  assumption of large $z$ is self-consistent. 

Translating back to the self energy via $\Sigma_R = z\sqrt{T/t}$, we obtain
\begin{equation}\Sigma_R \sim T \sqrt{\alpha \log(\alpha T t)}.\end{equation} 

The self energy apparently grows without bound at $z_0 = 0$ albeit extremely slowly. We interpret this unbounded growth as a symptom of the 
non-existence of the true critical point in $d=2$ and therefore the impossibility of a self-consistent treatment in this regime. 

\section{Conclusions}\label{conclu}
In this paper we have analyzed the superfluid quench, wherein an attractive interaction is suddenly turned on in a normal fluid of fermions. 
This interaction enhances superfluid fluctuations. There are two regimes: a disordered phase at weak interaction strength where the fluctuations saturate 
at a finite value; and the ordered phase, for strong interaction strength, where the fluctuations grow exponentially, 
leading eventually to spontaneous symmetry breaking. 

Between these two regimes is a dynamical critical point, where the fluctuations grow but order is not formed. We find that as with the usual equilibrium 
critical points, there is a notion of universality associated with this dynamical critical point. That is, once a small number of constants are fixed, 
the complete behavior of the superfluid fluctuations is determined by a function of the wavelength and time, with no further free parameters. 
The necessary parameters are the $r$ and $\ell$ given in Eq.~(\ref{eq:DREoM}).

Moreover, we find a signature of this universality in the lifetime of the fermions. The mechanism is essentially that the fermions near the Fermi energy 
scatter resonantly off of superfluid fluctuations. Thus the growing superfluid fluctuations lead to a singular feature in the fermion lifetime 
as a function of energy. We show that this singular feature inherits the universality of the dynamical critical point. In particular after fixing the Fermi 
velocity $v_F$ and normalized scattering rate $\alpha$, the energy and time dependence of the lifetime is completely determined.

The present work may be extended in several directions. One is the full development of the kinetic equation governing the fermion dynamics, 
to be published elsewhere. It would also be of interest to repeat this analysis for a disordered system to allow for comparison with 
pump-probe experiments.  Lastly, extending this treatment to include other fermion symmetry breaking channels, such as magnetic orders, 
or charge-density waves, would be fruitful.

We note that the perturbative calculation in $d=3$ gives a logarithmic correction $\sim \log t$ which grows large with $t$. 
This suggests that a dynamical RG conducted 
around the critical dimension $d=3$ may be a fruitful alternative way to approach this problem.

In this paper we have consider the fermions to initially be at finite temperature before the quench. 
A natural problem would be to consider the quench starting with fermions at zero temperature. This problem is more delicate for at least two reasons. 
Firstly the superfluid phase transition always occurs at finite temperature, therefore to approach the critical regime one would have to consider 
the temperature that is dynamically generated by the self-heating of the fermions. 
Secondly before the temperature is generated, the fermions are controlled by quantum fluctuations, leading to 
complex prethermal dynamics~\cite{Knap15}.
These difficulties aside, the problem appears deserving of future study.

{\sl Acknowledgements:}
This work was supported by the US National Science Foundation Grant NSF-DMR 1607059.

\begin{appendix}

\section{The $D$ propagator or Cooperon as correlators of Hubbard-Stratonovich fields}\label{ap1}
The final post-quench Hamiltonian is,
\begin{eqnarray}
H_f = H_i + \frac{u}{N}\sum_q \Delta^\dagger_q\Delta_q.
\end{eqnarray}
We will highlight the meaning of $D$ in an imaginary time formalism as the generalization
to real time Keldysh formalism is conceptually straightforward.

We may decouple the quartic interaction via a complex field $\phi_{q}$ for each momentum mode $q$,
\begin{eqnarray}
&&\prod_{q}e^{-\frac{u}{N}\Delta^\dagger_q\Delta_q}=
\int \biggl[\phi_{q},\phi^*_{q}\biggr]\nonumber\\
&&\times e^{-\frac{N}{u}|\phi_{q}|^2 + \phi_{q}\Delta^\dagger_q+\phi_q^*\Delta_q}.
\end{eqnarray}
In this picture, the action is quadratic in the fermionic fields. 
After integrating out the fermions one may write the partition function $Z$ as,
\begin{eqnarray}
&&Z=\int \biggl[\phi,\phi^*\biggr]
e^{-\frac{N}{u}\int dx |\phi|^2 +{\rm Tr}\ln\left[g^{-1}-\begin{pmatrix}0&\phi \\ \phi^* & 0 \end{pmatrix}\right]},
\end{eqnarray}
where $g^{-1}$ is the non-interacting fermionic Green's function in $2\times 2$ Nambu space. On expanding the ${\rm Tr}\ln$,
one obtains an action for the $\phi$ fields. Since the system is assumed to be in the normal phase, only
even powers of the $\phi$ field enter the action. 
Thus, we obtain,
\begin{eqnarray}
&&Z = \int \biggl[\phi,\phi^*\biggr]e^{-S\left(\hat{\phi}\right)}; \hat{\phi}= \begin{pmatrix}0&\phi \\ \phi^* & 0 \end{pmatrix},
\end{eqnarray}
where
\begin{eqnarray}
&&S= \frac{N}{u}\int dx |\phi(x)|^2 - \frac{1}{2}{\rm Tr}\biggl[g\hat{\phi}g{\hat \phi}\biggr] \nonumber\\
&&-\frac{1}{4}{\rm Tr}\biggl[g\hat{\phi}g{\hat \phi}g\hat{\phi}g{\hat \phi}\biggr].
\end{eqnarray}
The Gaussian approximation involves keeping only quadratic terms in the $\phi$ fields. The coefficient of
$\phi^2$ in the second term
in the action is recognized as the polarization bubble $\Pi \equiv gg$.
The equation of motion at Gaussian order is,
\begin{eqnarray}
\biggl[\frac{1}{u}- \Pi\biggr]D = 1 \label{Lang1},
\end{eqnarray}
where trace over the fermions gives an additional factor of $N$. In the next sub-section
we show that Eq.~\eqref{Lang1} is equivalent to a classical Langevin equation for the Hubbard-Stratonivich fields
when the fermions are at non-zero temperatures.

\section{Properties of the $\Pi$ and Relationship to Model-A}\label{ap2}
The fermionic distribution function before the quench is,
$n_{\sigma}(k)=1/\left(e^{\frac{\xi_k}{T}}+1\right),\xi_k=\epsilon_k-\mu$. We measure all energies
relative to the chemical potential. 

The Keldysh component of the polarization bubble is found to be,
\begin{eqnarray}
&&i\Pi^K(q,t,t')=\sum_k e^{-i\left(\xi_{k\uparrow}+\xi_{-k+q\downarrow}\right)(t-t')}\nonumber\\
&&\times \biggl[n_{\sigma}(k)n_{-\sigma}(-k+q)\nonumber\\+
&&\left(1-n_{\sigma}(k)\right)\left(1-n_{-\sigma}(-k+q)\right)
\biggr],\label{pik1}
\end{eqnarray}
while the retarded component is
\begin{eqnarray}
&&i\Pi^R(q,t,t')
=\theta(t-t')\sum_k e^{-i\left(\xi_{k\uparrow}+\xi_{-k+q\downarrow}\right)(t-t')}\nonumber\\
&&\times \biggl[-n_{\sigma}(k)- n_{-\sigma}(-k+q)+1\biggr].\label{pir1a}
\end{eqnarray}

Since the $\Pi$ are time-translation invariant in this approximation, it is helpful to write them
in frequency space,
\begin{eqnarray}
&&\Pi^R(q,\omega)=-\frac{1}{2}\sum_k \frac{\tanh\biggl[\frac{\xi_k}{2T}\biggr] + \tanh\biggl[\frac{\xi_{k-q}}{2T}\biggr]}
{\omega - \xi_{k}-\xi_{k-q}+i\delta},\\
&&\Pi^K(q,\omega)=2i\pi\sum_k \biggl(n\biggl[\frac{\xi_k}{T}\biggr]n\biggl[\frac{\xi_{-k+q}}{T}\biggr] \nonumber\\
&&+
\biggl(1-n\biggl[\frac{\xi_k}{T}\biggr]\biggr)\biggl(1-n\biggl[\frac{\xi_{-k+q}}{T}\biggr]\biggr)
\biggr)
\delta\left(\omega - \xi_{k}-\xi_{k-q}\right).\nonumber\\
\end{eqnarray}
Now we use the fact that $1-2n(x) = \tanh(x/2)$ and using that $\coth(a) \coth(b) +1 = \coth(a+b)(\coth(a)+\coth(b))$,
one may show that fluctuation dissipation theorem (FDT) is obeyed,
\begin{eqnarray}
\Pi_K(q,\omega) =\coth\biggl(\frac{\omega}{2T}\biggr)\biggl[\Pi_R(q,\omega)-\Pi_A(q.\omega)\biggr].
\end{eqnarray}
It should be emphasized that this FDT is simply inherited from the properties of the initial state. In a better approximation, 
the FDT will cease to hold as the system goes through the process of thermalization. 
 
We are interested in the dynamics of the soft Cooperon mode, which evolves on a timescale much larger than $T^{-1}$. Therefore we expand $\Pi^R\!\left(q,\omega\right)$ 
in  $\omega/T$, $q^2/T$. The constant term,
\begin{align}
\Pi^R\!\left(0,0\right) &= 
	\sum_k\frac{\tanh\left[\frac{\xi_k}{2T}\right]}{2\xi_k -i\delta}\nonumber\\
	&\approx \frac{1}{2}\nu \log E_F /T, 
\end{align}
is the usual Cooper logarithm, where $\nu$ is the density of states and $E_F$ is some bandwidth or Fermi energy. The coefficient of $\omega/T$ is
\begin{align}
\frac{\partial}{\partial \omega} \Pi^R\!\left(0,0\right) 
&= \sum_k\frac{ \tanh\left[ \frac{\xi_k}{2T}\right] }{\left(2\xi_k - i\delta\right)^2}\nonumber\\
&= \pi i \sum_k \tanh\left[ \frac{\xi_k}{2T}\right] \delta'\!\left(2\xi_k\right)\nonumber\\
&= \frac{i \nu \pi}{2 T}
\end{align}
which is purely imaginary in the absence of particle hole asymmetry. 

For the coefficient of $q^2/T$, we expand the dispersion as $ \epsilon_{k-q} = \epsilon_k - \vec{q}\cdot \vec{v}_k$ and obtain,
\begin{align}
\frac{\partial^2}{\partial q^2} \Pi^R\!\left(0,0\right) 
&= \sum_k \frac{\tanh''\left[\frac{\xi_k}{2T}\right] (\vec{v}_k/(2T))^2 }
{2\xi_k-i\delta}\nonumber\\
&= \frac{\nu}{8 T^2}\langle v^2_k\rangle_{FS} \int_{-\infty}^\infty dx \frac{\tanh'' x}{x},
\end{align}
where $\langle \cdot\rangle _{FS}$ is the average over the Fermi surface and the integral evaluates to the constant 
$28\zeta(3)/\pi^2 \approx 3.41$.

With this expansion for $\Pi_R$, the FDT gives that $\Pi_K$ is given by 
\begin{align}
i\Pi_K(\omega, q = 0) &\sim 2\nu\pi 
\end{align}
and therefore in real time 
\begin{equation}
i\Pi_K(t,t') = \nu \delta(t-t'), 
\end{equation}
as discussed in the text. The fact that $\Pi_K$ is well approximated by a delta function is entirely a consequence of the fact that we are interested 
in timescales much longer than $T^{-1}$, because the Cooperon dynamics are governed by much longer timescales at the critical point.

Thus in summary, the above behavior for $\Pi$ together with how it affects the equation of motion of $D$ (Eq.~\eqref{Lang1}) show that the Cooperon
obeys model-A dynamics close to the critical point.

\section{Interacting Cooperons in the Hartree-Fock Approximation}\label{ap3}

For the sake of completeness we outline how the results for interacting bosons used in the main text were obtained.
We employ a Hartree-Fock approach, although the same scaling forms can be obtained with an
$\epsilon$-expansion~\cite{Janssen1988}. The Hartree-Fock approximation for the bosons is justified as the $N\rightarrow \infty$
limit of a bosonic model where $N$ denotes the number of components of the boson field. This $N$ should not be confused with the
orbital index of the fermions used in the main text.
 
The Hartree-Fock equations of motion are,
\begin{eqnarray}
&&\partial_t D_R(k,t,t') + \left[k^2 + r_{\rm eff}(t)\right]D_R(k,t,t')=-\delta(t-t'), \nonumber\\
&&\Rightarrow D_R(k,t,t') = -\theta(t-t')e^{-k^2(t-t')}  e^{-\int_{t'}^t dt_1r_{\rm eff}(t_1)},
\end{eqnarray}
where the mass obeys the equation of motion
\begin{eqnarray}
&&r_{\rm eff}(t) = r + u \int \frac{d^dq}{(2\pi)^d}iD_K(q,t,t),\\
&&D_K = D_R \circ \Pi_K \circ D_A.
\end{eqnarray}
The overdamped dynamics of $D_R$ is entirely due to the underlying
finite temperature Fermi sea which gives $u^{-1}-\Pi^{R}= r +i\omega$.

If we employ the Gaussian expression for $D_K(q,t,t) \rightarrow \frac{T}{q^2+r}\left[1-e^{-2(q^2+r)t}\right]$,
we find that
\begin{eqnarray}
r_{\rm eff}(t)-r_c \rightarrow \int q^{d-1}dq \frac{1}{q^2}e^{-2 q^2 t} \propto \frac{1}{t^{\frac{d}{2}-1}}.
\end{eqnarray}
The above shows that scaling emerges only if we set $d=4$ in the above Gaussian result, showing that
the upper critical dimension of the theory is $d=4$. 

Thus with the ansatz,
\begin{eqnarray}
r_{\rm eff}(t) =-\frac{a}{t},
\end{eqnarray}
we obtain,
\begin{eqnarray}
D_R(q,t,t')= -e^{-q^2(t-t')}\left(\frac{t}{t'}\right)^a,
\end{eqnarray}

For $D_K$ we have,
\begin{eqnarray}
&&iD_K(q,t,t'; t>t') \nonumber\\
&&= 2T \int_0^{t'}dt_1e^{-q^2(t-t_1)-q^2(t'-t_1)}\left(t/t_1\right)^a \left(t'/t_1\right)^a,\nonumber\\
&&= 2T e^{-q^2(t+t')}(tt')^a\int_0^{t'}dt_1 e^{2 q^2t_1} t_1^{-2a}.\label{Dint}
\end{eqnarray}
For $q^2 t' \ll 1$, we obtain aging behavior,
\begin{eqnarray}
iD_K(q,t, t';q^2 t' \ll 1)= c e^{-q^2 t}t^a (t')^{1-a},
\end{eqnarray}

For equal times we may write,
\begin{eqnarray}
&&i D_K(q,t,t)= 2Te^{-2 q^2 t}t^{2a}\int_0^{t}dt_1 e^{2 q^2t_1}t_1^{-2a}, \nonumber\\
&&= \frac{T}{q^2}F(2 q^2t),\nonumber\\
&& F(x) = e^{-x}x^{2a}\int_0^x dy' e^{y'}y'^{-2a}, \nonumber\\
&&= x \int_0^1 dy e^{-xy}(1-y)^{-2a}.\label{FKdef}
\end{eqnarray}
Note that $F(x=0)=0$ and $F(x=\infty)=1$.

In order to solve for $a$, we use that
\begin{eqnarray}
&&r_{\rm eff}(t) = r_{\rm eff}(\infty) \nonumber\\
&&+ u \int_q\biggl[iD_K(q,t,t)-iD_K(q,\infty,\infty)\biggr].
\end{eqnarray}
At criticality $r_{\rm eff}(\infty)=0$ and $iD_K(q,\infty,\infty)= T/q^2$. Using this,
\begin{eqnarray}
r_{\rm eff}(t) = uA_d \int_0^{\Lambda} d q q^{d-1}\frac{1}{q^2}\biggl[F(2 q^2 t)-1\biggr].
\end{eqnarray}
where $A_d=$ the surface area of a $d$-dimensional unit sphere.

The above may be recast as
\begin{eqnarray}
-\frac{a}{t}=\frac{uA_d}{t^{-1+d/2}}\int_0^{2\Lambda^2 t} dx x^{-2 + d/2}\biggl[F(x)-1\biggr].
\end{eqnarray}
Thus we may write, defining $\epsilon=4-d$,
\begin{eqnarray}
&&a= - u A_dt^{\epsilon/2}\biggl\{\int_0^{\infty}dx x^{-\epsilon/2}\left[F(x)-1\right]\nonumber\\
&&-\int_{2\Lambda^2 t}^{\infty}dx x^{-\epsilon/2}\left[F(x)-1\right]\biggr\}.
\end{eqnarray}
The first integral above increases in time as $t^{\epsilon/2}$ unless
\begin{eqnarray}
\int_0^{\infty}dx x^{-\epsilon/2}\left[F(x)-1\right]= 0\label{cond1}.
\end{eqnarray}
Notice that in Eq.~\eqref{FKdef}, to avoid infra-red singularity $2a < 1$. Then,
\begin{eqnarray}
&&F(x, a<1/2) = e^{-x} x (-x)^{2 a-1} \nonumber\\
&&\times \biggl[\Gamma (1-2 a)-\Gamma (1-2 a,-x)\biggr]\label{cond2}.
\end{eqnarray}
Since $F(0)=0$, we require $\epsilon/2 <1$ or $d>2$ to make the $\int dx x^{-\epsilon/2}$ infra-red convergent.  
Substituting Eq.~\eqref{cond2} in Eq.~\eqref{cond1}, we obtain
\begin{eqnarray}
&&\int_0^{\infty}dx x^{-\epsilon/2}\left[F(x)-1\right]_{\epsilon<2},\nonumber\\
&&=  -\frac{\Gamma(1-2a)\Gamma(\epsilon/2)\Gamma(1-\epsilon/2)}{\Gamma(-2a +\epsilon/2)}=0\nonumber\\
&&\Rightarrow a=\epsilon/4.
\end{eqnarray}
Thus we have derived the quoted scaling forms, and also the initial slip exponent $a=\theta=\epsilon/4$.

It is also useful to note that for $t>t'$ but general $q^2t,q^2t'$, one obtains from Eq.~\eqref{Dint},
\begin{eqnarray}
iD_K(q,t,t') = \frac{T}{q^2}e^{-q^2(t-t')}\left(t/t'\right)^{\theta}F(2q^2t')
\end{eqnarray}
Defining
\begin{eqnarray}
F_K(x) = F(x)/x
\end{eqnarray}
\begin{eqnarray}
iD_K(q,t,t') = Tt'e^{-q^2(t-t')}\left(t/t'\right)^{\theta}F_K(2q^2t')
\end{eqnarray}

\end{appendix}

%\bibliography{quench}

\end{document}